\documentclass[preprint,showpacs,preprintnumbers,amsmath,amssymb,nofootinbib]{revtex4}
\usepackage{graphicx,bm} 
\usepackage{dcolumn}    

\begin{document}
\draft
\title{Ferromagnetism of two-flavor quark matter in chiral and/or 
color-superconducting phases at zero and finite temperatures}
 
\author{M. Inui, H. Kohyama, A. Ni\'{e}gawa}
\address{Graduate School of Science, Osaka City University, 
Sumiyoshi-ku, Osaka 558-8585, JAPAN}
\date{Received today}
\begin{abstract}

We study the phase structure of the unpolarized and polarized two-flavor quark 
matters at zero and finite temperatures within the 
Nambu--Jona-Lasinio (NJL) model. We focus on the region, which 
includes the coexisting phase of quark-antiquark and diquark 
condensates. Generalizing the NJL model so as to describe the 
polarized quark matter, we compute the thermodynamic potential as a 
function of the quark chemical potential ($\mu$), the temperature ($T$), and 
the polarization parameter. The result heavily depends on 
the ratio $G_D / G_S$, where $G_S$ is the quark-antiquark coupling 
constant and $G_D$ is the diquark coupling constant. We find that, 
for small $G_D / G_S$, the $\lq\lq$ferromagnetic" phase is energetically favored 
over the $\lq\lq$paramagnetic" phase. 
On the other hand, for large $G_D / G_S$, 
there appears the window in the ($\mu, T$)-plane, in which the 
$\lq\lq$paramagnetic" phase is favored. 
\end{abstract} 
\maketitle
\setcounter{section}{0}
\setcounter{equation}{0}
\def\theequation{\mbox{\arabic{section}.\arabic{equation}}} 
\section{Introduction} 
Quantum chromodynamics (QCD) is the fundamental theory of strong 
interaction between quarks and gluons. QCD is an asymptotically 
free theory and perturbation theory may be used at high density 
and/or temperature. One-gluon exchange interaction between two 
quarks in the color antitriplet channel is attractive. Then one 
expects that a quark matter undergoes a phase transition to color 
superconducting phases at high density and relatively low 
temperature. Around two decades ago, it was shown that such phase 
transitions really take place at high density and low 
temperature \cite{fra}. Around a decade ago, it was found 
\cite{rapp} that the color superconducting gaps are of $O$(100MeV), 
which are comparable with nuclear matter density. 

Quark matter is expected to be produced in heavy-ion-collision 
experiments, and might exist inside compact stars. If quark stars 
exist, they are quark matters in itself. Such quark matters are 
the system with intermediate density at low temperature, so that 
perturbative QCD is not applicable. Although lattice QCD is powerful 
for studying the system with vanishing net baryon density, it is not 
efficient, at the present moment, for studying nonzero baryon 
density systems due to the \lq\lq sign problem''. Then, for studying 
such systems, one should rely on some effective theories of QCD, 
such as the extended Nambu--Jona-Lasinio (NJL) model \cite{NJL}. 
The NJL model incorporates above-mentioned attractive nature of the 
one-gluon exchange interaction between two quarks in the color 
antitriplet channel. Furthermore, it successfully describes static 
properties of the pion (e.g., pion mass and decay constant) in the 
hadoronic phase through chiral phase transition (see, e.g., 
\cite{HK}). 

Theoretical studies using the extended NJL model as well as other 
approaches have disclosed possible existence of various color 
superconducting phases; regular 2SC phase, charge-neutral gapless 
2SC phase, color-flavor-locked phase, and so on. Furthermore, 
it has been found that, at moderate baryon density region, the 
chiral and diquark condensates co-exist \cite{coex}. (For recent reviews, see, 
e.g., \cite{rischke,mei}.) 

Magnetic property of quark matter is yet another important issue. 
Tatsumi was the first who pointed out the possible instability of 
a quark matter against the spin polarization (ferromagnetism) 
\cite{tatsumi}: Within the one-gluon-exchange approximation in QCD, 
he computed the energy density ${\cal E}$ of quark matter at zero 
temperature as a function of polarization parameter $p$, and found 
that, at relatively low density, ${\cal E} (p)$ decreases as $p$ 
increases. Since then, there appear quite a few papers that are 
devoted to possible ferromagnetism in quark matter \cite{FM}. 

In this paper, we generalize the Tatsumi's analysis \cite{tatsumi} 
to the case of polarized two-flavor quark matters with moderate 
baryon density at zero and finite temperatures. We introduce two 
types of polarized quark matters; the one is the \lq\lq 
magnetic-moment--polarized'' quark matters and the other is the 
\lq\lq spin-polarized'' quark matters. For dealing with such 
systems, we generalize the extended NJL model and employ the 
mean-field or Hartree approximation. 

In Sec. II, we briefly describe the extended NJL model and 
generalize it so as to deal with polarized quark matters. 
In Sec. III, we derive the thermodynamic potential $\Omega$ for the 
two types of polarized quark matters at finite temperature within 
the mean field approximation. (Concrete derivation is given in 
Appendix D.) We find that the thermodynamic potentials for two types 
of polarized quark matters are of the same form. Difference may 
arise when the residual interactions are taken into account. In Sec. 
IV, we present the results obtained through numerical analysis. As 
in various analyses within the extended NJL model, the results are 
sensitive to the quark-antiquark coupling constant $G_S$ and the 
diquark coupling constant $G_D$. We show that, in the case of $G_D / 
G_S \lesssim 1.15$, the polarized state is energetically favored. While, for $G_D 
/ G_S \gtrsim 1.15$, there appears the window in a $(\mu, T)$-plane, 
in which the unpolarized state is favored, where $\mu$ is the quark 
chemical potential and $T$ is the temperature. Sec. V is devoted to 
summary and conclusion. In Appendix A, we briefly review the role of 
the projectors ${\bf P}_s^{(\tau)}$ ($\tau = \pm$, $s = \pm$), Eq. 
(\ref{pro}). In Appendix B, we show that the polarized quark numbers 
are conserved to the first order of coupling constants. In Appendix 
C, for completeness, we give the forms of the quark propagators in 
the polarized quark matters. 
\setcounter{section}{1}
\setcounter{equation}{0}
\def\theequation{\mbox{\arabic{section}.\arabic{equation}}} 
\section{Extended Nambu--Jona-Lasinio model and its generalization} 
\subsection{Extended Nambu--Jona-Lasinio model} 
For describing the two-flavor quark matters, we adopt the extended 
Nambu--Jona-Lasinio model with the scalar-, pseudoscalar-, and 
scalar diquark-channels taken into account, whose lagrangian 
density reads \cite{mei} 
\[
{\cal L} = \bar{q} \left( i 
\partial\kern-0.em\raise0.17ex\llap{/}\kern0.15em\relax - m_0 
\right) q + G_S \left[ \left( \bar{q} q \right)^2 + \left( \bar{q} i 
\gamma_5 \vec{\tau} q \right)^2 \right] + G_D \left[ \left(i 
\bar{q}^C \epsilon \epsilon^b \gamma_5 q \right) \left(i \bar{q} 
\epsilon \epsilon^b \gamma_5 q^C \right) \right] \, , 
\] 
where $m_0$ is the current quark mass, $q^C = C \bar{q}^T$, 
$\bar{q}^C = q^T C$ with $C = i \gamma^2 \gamma^0$ the 
charge-conjugation matrix. The quark field is a doublet in a flavor 
space and a triplet in a color space, $q \equiv q_{i \alpha}$ with 
$i = 1, 2$ and $\alpha =$ r(ed), g(reen), b(lue). The Pauli matrices 
$\vec{\tau} = (\tau^1, \tau^2, \tau^3)$ and $\epsilon \equiv i 
\tau^2$ act on the flavor space, while $(\epsilon^b)^{\alpha \beta} 
\equiv \epsilon^{\alpha \beta b}$ $(\epsilon^{r g b} = 1)$ is a 
totally antisymmetric tensor in a color space. Although the coupling 
constants, $G_S$ and $G_D$, enjoy the relation $G_D / G_S = 3 / 4$, 
we regard, as in \cite{mei}, $G_D / G_S$ as a free parameter. 
\subsection{Generalization to the case of polarized quark 
matter} 
For dealing with polarized quark matters, we first introduce a 
spin-polarization vector $n^\mu (\vec{p})$, which is obtained from 
its rest-frame form, $n^\mu (\vec{0}) = (0, \vec{e}^{\, z}) = (0, 0, 
0, 1)$, through a Lorentz transformation \cite{tatsumi}, 
\begin{equation} 
n^0 (\vec{p}) = \frac{\vec{p} \cdot \vec{e}^{\, z}}{m_0} \, , 
\;\;\;\;\;\; \vec{n} (\vec{p}) = \vec{e}^{\, z} + 
\frac{(\vec{p} \cdot \vec{e}^{\, z}) \, \vec{p}}{m_0 (E_p + m_0)} 
\, . 
\label{2.1}
\end{equation} 
Here $E_p$ $(= \sqrt{p^2 + m^2_0})$ is the energy of the 
(anti)quark with momentum $\vec{p}$. Obviously, we have the 
relations, 
\begin{equation} 
n^2 (\vec{p}) \equiv n_0^2 (\vec{p}) - \vec{n}^{\, 2} (\vec{p}) = 
- 1 \, , \;\;\;\;\; E_p n_0 (\vec{p}) = \vec{p} \cdot \vec{n} 
(\vec{p}) \, . 
\label{rela} 
\end{equation} 
Using Eq. (\ref{2.1}), we introduce a set of projectors onto the 
spin-polarized states: 
\begin{equation} 
{\cal P}_{\pm} (\vec{p}) = \frac{1}{2} \left( 1 \pm \gamma_5 
n\kern-0.em\raise0.17ex\llap{/}\kern0.15em\relax (\vec{p}) 
\right) \, . 
\label{spinpro}
\end{equation} 

We also introduce a set of energy projectors \cite{mei}: 
\begin{equation} 
\tilde{\Lambda}_\pm (\vec{p}) = \frac{1}{2 E_p} \left[ E_p \pm 
\gamma_0 \left( \vec{\gamma} \cdot \vec{p} - m_0 \right) \right] 
\, . 
\label{energypro}
\end{equation} 
It should be noted that $\gamma_0 \tilde{\Lambda}_- (\vec{p}) = (m_0 
/ E_p) \Lambda_+ (\vec{p})$ and $\gamma_0 \tilde{\Lambda}_+ 
(- \vec{p}) = - (m_0 / E_p) \Lambda_- (\vec{p})$, where 
$\Lambda_{+ (-)} (\vec{p})$ is the projector onto the positive 
(negative) energy state. Now we introduce a set of projectors: 
\begin{eqnarray}
&& {\bf P}^{(+)}_s (\vec{p}) \equiv {\cal P}_s (\vec{p}) 
\tilde{\Lambda}_+ (\vec{p}) = \tilde{\Lambda}_+ (\vec{p})   
{\cal P}_s (- \vec{p}) \nonumber \\ 
&& {\bf P}^{(-)}_s (\vec{p}) \equiv {\cal P}_s (- \vec{p}) 
\tilde{\Lambda}_- (\vec{p}) = \tilde{\Lambda}_- (\vec{p})   
{\cal P}_s (\vec{p}) \, , 
\label{pro} \\ 
&& {\bf P}^{(\tau)}_s (\vec{p}) {\bf P}^{(\tau')}_{s'} (\vec{p}) 
= \delta^{\tau \tau'} \delta_{s s'} {\bf P}^{(\tau)}_s (\vec{p}) 
\, , \;\;\;\;\;\;\; (\tau, \tau' = \pm, \; s, s' = \pm) \, . 
\label{keypro}
\end{eqnarray}
In Appendix A, we show that, when acting on the quark field $q (x)$, 
${\bf P}_\pm^{(-)} (i \nabla)$ projects out onto the \lq\lq 
spin-up'' (\lq\lq spin-down'') positive energy-state, while 
${\bf P}_\pm^{(+)} (i \nabla)$ projects out onto the \lq\lq 
spin-up'' (\lq\lq spin-down'') negative energy-state. 

The projectors ${\bf P}_s^{(\pm)} (\vec{p})$ enjoy the following 
relations: 
\begin{eqnarray}
\gamma_0 {\bf P}_s^{(\pm)} (\vec{p}) \gamma_0 & = & 
{\bf P}_s^{(\pm)} (- \vec{p}) \, , 
\label{D1} \\ 
\gamma_5 {\bf P}_s^{(\pm)} (\vec{p}) \gamma_5 & = & 
{\bf P}_{- s}^{(\mp)} (- \vec{p}) \, , 
\label{D2} \\ 
C \left( {\bf P}_s^{(\pm)} (\vec{p}) \right)^T C & = & - 
{\bf P}_s^{(\mp)} (\vec{p}) \, . 
\label{D3} 
\end{eqnarray}

We are now in a position to describe the \lq\lq polarized'' quark 
matter. From the above observation, we introduce 
the \lq\lq spin up''/\lq\lq spin down'' quark number, $Q_\pm^{(-)}$, 
and the \lq\lq spin up''/\lq\lq spin down'' antiquark number, 
$Q_\pm^{(+)}$, where 
\[
Q_s^{(\tau)} = \int d^3 x \, {\cal Q}_s^{(\tau)} \equiv \int d^3 x 
\, q^\dagger (x) {\bf P}_s^{(\tau)} (i \nabla) q (x) \;\;\;\;\;\; 
(\tau = \pm, \; s = \pm) \, . 
\]
\subsubsection{Magnetic-moment-polarized quark matter}
Since a \lq\lq spin up'' (\lq\lq spin down'') quark and a  \lq\lq 
spin down'' (\lq\lq spin up'') antiquark feel the same 
electromagnetic force, we introduce two quark chemical potentials: 
$\mu_+$ is conjugate to the net quark-number charge with \lq\lq positive 
magnetic moment (MM)'', 
\begin{equation}
Q_+ \equiv Q_+^{(-)} + Q_-^{(+)} \; 
\left( \equiv \int d^3 x \, {\cal Q}_+ \right) \, , 
\label{tasu}
\end{equation}
and $\mu_-$ is conjugate to the net quark-number charge with \lq\lq 
negative MM'', 
\begin{equation}
Q_- \equiv Q_-^{(-)} + Q_+^{(+)} \; 
\left( \equiv \int d^3 x \, {\cal Q}_- \right) \, . 
\label{hiku}
\end{equation}
It can readily be shown that $\left[Q_+, Q_- \right] = 0$. We call 
the quark matter with $\mu_+ \neq \mu_-$ the \lq\lq MM-polarized'' 
quark matter. For dealing with such quark matters, the term that 
depends on $\mu_+$ and $\mu_-$ should be added to the Lagrangian 
density; 
\begin{equation} 
\tilde{\cal L} \rightarrow \tilde{\cal L} + \mu_+ {\cal Q}_+ + 
\mu_- {\cal Q}_- \, . 
\label{hajime}
\end{equation} 

The polarized charges $Q_+$ and $Q_-$ are not conserved. We show in 
Appendix B that they are conserved up to and including the first 
order of coupling constants $G_S$ and $G_D$. It is to be noted in 
passing that, when the mean-field approximation is employed, even 
the unpolarized charge $Q_+ + Q_-$ turns out not to be 
conserved\footnote{We thank A. Oguri for discussion on this point.}. 
\subsubsection{Spin-polarized quark matter}
We define \lq\lq spin-polarized'' quark matters by introducing two 
chemical potentials, $\mu_+$ and $\mu_-$: $\mu_+$ is conjugate to 
$Q_+^{(-)} + Q_+^{(+)}$ $(\equiv Q_+')$ and $\mu_-$ is conjugate to 
$Q_-^{(-)} + Q_-^{(+)}$ $(\equiv Q_-')$. One can show that $\left[ 
Q_+', Q_-' \right] = 0$. In this case, the counterpart to Eq. 
(\ref{hajime}) is 
\begin{equation} 
\tilde{\cal L} \rightarrow \tilde{\cal L} + \mu_+ {\cal Q}_+' + 
\mu_- {\cal Q}_-' \, . 
\label{hajime1}
\end{equation} 
We see in Appendix B that the charges $Q_+'$ and $Q_-'$ are 
conserved up to and including $O (G_S, G_D)$. 
\subsection{Mean field approximation} 
We employ the mean field approximation to get 
\begin{eqnarray}
\tilde{\cal L} & = & \bar{q} \left( i 
\partial\kern-0.em\raise0.17ex\llap{/}\kern0.15em\relax - m 
\right) q - \frac{1}{2} \Delta^{*b} \left(i \bar{q}^C \epsilon 
\epsilon^b \gamma_5 q \right) - \frac{1}{2} \Delta^b \left(i 
\bar{q} \epsilon \epsilon^b \gamma_5 q^C \right) \nonumber \\ 
&& - \frac{\sigma^2}{4 G_S} - \frac{\Delta^{*b} \Delta^b}{4 G_D} 
\, , 
\label{heikin}
\end{eqnarray}
where $\langle \bar{q} i \gamma_5 \vec{\tau} q \rangle = 0$ has been 
assumed \cite{mei} and 
\begin{eqnarray}
m & = & m_0 + \sigma = m_0 - 2 G_S \langle \bar{\psi} \psi \rangle 
\, , 
\label{m} \\ 
\Delta^b &=& - 2 G_D \langle i \bar{q}^C \epsilon \epsilon^b 
\gamma_5 q \rangle \, , \;\;\;\;\;\;\; \Delta^{*b} = - 2 G_D 
\langle i \bar{q} \epsilon \epsilon^b \gamma_5 q^C \rangle \, . 
\end{eqnarray}
$m$ in Eq. (\ref{m}) is the constituent quark mass, which has 
received the contribution $\sigma$ from the chiral condensate, if 
any. Then, when we use in the sequel various quantities defined in the last 
subsection, we replace the current quark mass 
$m_0$ with $m$. 

From Eq. (\ref{heikin}), we see that the red and green quarks 
participate in the diquark condensate, while the blue quarks do not. 
\setcounter{section}{2}
\setcounter{equation}{0}
\def\theequation{\mbox{\arabic{section}.\arabic{equation}}} 
\section{Thermodynamic potential in the mean-field approximation} 
\subsection{Magnetic-moment-polarized quark matter} 
Throughout in the sequel of this paper, we 
use the imaginary-time formalism. The grand partition function is 
defined by 
\begin{eqnarray}
{\cal Z} & = & \int {\cal D} \bar{q} {\cal D} q \exp \left\{ 
\int_0^\beta d \tau \int d^{\, 3} x \left( \tilde{\cal L}_E + 
\sum_{s = \pm} \mu_s {\cal Q}_s \right) \right\} \, , 
\label{part} 
\end{eqnarray}
where $\beta = 1 / T$ is the inverse temperature, $\tau = i x^0$ ($0 
< \tau< \beta$) is the Euclidean time, and $\tilde{\cal L}_E = 
\tilde{\cal L} \, \rule[-1.5mm]{.10mm}{6.0mm} 
\raisebox{-1.5mm}{\scriptsize{$\; x_0 = - i \tau$}}$. 

Since the exponent of Eq. (\ref{part}) is bilinear in fields, 
${\cal Z}$ is factorized as 
\begin{equation}
{\cal Z} = {\cal Z}_{\mbox{\scriptsize{const}}} {\cal Z}_b 
{\cal Z}_{r, g} \, . 
\label{zeroth}
\end{equation}
${\cal Z}_{\mbox{\scriptsize{const}}}$ reads 
\begin{equation}
\ln {\cal Z}_{\mbox{\scriptsize{const}}} = - V \beta \left( 
\frac{\sigma^2}{4 G_S} + \frac{\Delta^{b *} \Delta^b}{4 G_D} \right) 
\, , 
\label{arya}
\end{equation}
where $V$ is the volume of the system. The contribution from the blue 
quarks is 
\begin{eqnarray}
{\cal Z}_b = \int {\cal D} \bar{q}_b {\cal D} q_b \exp \left\{ 
\int_0^\beta d \tau \int d^{\, 3} x \left( \frac{1}{2} \bar{q}_b 
\left( G_0^+ \right)^{- 1} q_b + \frac{1}{2} \bar{q}_b^C \left( 
G_0^- \right)^{- 1} q_b^C \right) \right\} \, , 
\label{blue} 
\end{eqnarray}
where $[ G_0^- (i \partial) ]^{- 1} = - C \left( [ G_0^+ (- i 
\partial)]^{- 1} \right)^T C$ and then 
\begin{equation}
\left( G_0^\pm \right)^{- 1} = i 
\partial\kern-0.em\raise0.17ex\llap{/}\kern0.15em\relax - m \pm 
\gamma_0 \sum_{\tau, \, 
s = \pm} \mu_{\mp \tau s} {\bf P}_s^{(\tau)} (i \nabla) \, . 
\label{kaku1}
\end{equation}
For the contribution from the red and green quarks, $Q = q_{r,g}$, 
we have 
\begin{eqnarray}
{\cal Z}_{r, \, g} & = & \int {\cal D} \bar{Q} {\cal D} Q \exp 
\left\{ \int_0^\beta d \tau \int d^{\, 3} x \left[ \frac{1}{2} 
\bar{Q} \left( G_0^+ \right)^{- 1} Q \right. \right. \nonumber \\ 
&& \left. \left. + \frac{1}{2} \bar{Q}^C \left( G_0^- \right)^{- 1} 
Q^C + \frac{1}{2} \bar{Q} \Delta^- Q^C + \frac{1}{2} \bar{Q}^C 
\Delta^+ Q \right] \right\} \, , 
\label{redgreen} 
\end{eqnarray}
where 
\begin{equation}
\Delta^- = - i \Delta \epsilon \epsilon^b \gamma_5 \, , \;\;\;\;\;
\Delta^+ = - i \Delta^* \epsilon \epsilon^b \gamma_5 \, . 
\label{Delta}
\end{equation}

We employ the Nambu-Gorkov formalism through introducing 
\begin{eqnarray*}
\Psi_b & = & \left( 
\begin{array}{c} 
q_b \\ 
q_b^C 
\end{array} 
\right) \, , \;\;\; \bar{\Psi}_b = \left( \bar{q}_b \;\, \bar{q}_b^C 
\right) \, , \\ 
\Psi & = & \left( 
\begin{array}{c} 
Q \\ 
Q^C 
\end{array} 
\right) \, , \;\;\; \bar{\Psi} = \left( \bar{Q} \;\, \bar{Q}^C 
\right) \, . 
\end{eqnarray*}
Going into a momentum space, we have (cf. Eq. (\ref{kaku1})) 
\begin{equation} 
\left( G_0^\pm \right)^{- 1} = \sum_{\tau, \, s = \pm} \left( p_0 + 
\tau E_p \pm \mu_{\mp \tau s} \right) \gamma_0 {\bf P}_s^{(\tau)} 
(- \vec{p}) \, , 
\label{kakuyo} 
\end{equation}
where 
\begin{equation} 
p_0 = i p_{0 E} = \frac{i \pi}{\beta} (2 n + 1) \;\;\;\;\;\;\;\;\: 
(n = ..., -2, -1, 0, 1, 2, ...) \, . 
\label{matsu}
\end{equation} 
Eqs. (\ref{blue}) and (\ref{redgreen}) turn out, in respective 
order, to 
\begin{eqnarray*}
{\cal Z}_b & = & \int {\cal D} \Psi_b \exp \left\{ \frac{1}{2} 
\sum_{n, \, \vec{p}} \bar{\Psi}_b \left( \beta G_0^{-1} \right) 
\Psi_b \right\} = \left[ \mbox{Det} \left( - \beta G_0^{- 1} \right) 
\right]^{1 / 2} \, , \\ 
{\cal Z}_{r, g} & = & \int {\cal D} \Psi \exp \left\{ \frac{1}{2} 
\sum_{n, \, \vec{p}} \bar{\Psi} \left( \beta G^{-1} \right) \Psi 
\right\} = \left[ \mbox{Det} \left( - \beta G^{- 1} \right) 
\right]^{1 / 2} \, .  
\end{eqnarray*}
Here 
\begin{equation} 
G_0^{- 1} = \left( 
\begin{array}{cc} 
\left( G_0^+ \right)^{- 1} {\bf 1}_f \; & \, 0 \\ 
0 \; & \, \left( G_0^- \right)^{- 1} {\bf 1}_f 
\end{array} 
\right) \, , \;\;\;\;\; 
G^{- 1} = \left( 
\begin{array}{cc} 
\left( G_0^+ \right)^{- 1} {\bf 1}_f {\bf 1}_c^\perp \; & \, 
\Delta^- \\ \Delta^+ \; & \, \left( G_0^+ \right)^{- 1} {\bf 1}_f 
{\bf 1}_c^\perp 
\end{array} 
\right) \, , 
\label{matri}
\end{equation}
where $({\bf 1}_f)^{ij} = \delta^{ij}$ in the flavor space, and 
$({\bf 1}_c^\perp)^{\alpha \beta} = \delta^{\alpha \beta} - 
\delta^{\alpha b} \delta^{\beta b}$ in the color space. 

Although the forms of the propagators, $G_0$ and $G$, are not 
necessary for our purpose, we give them in Appendix C for 
completeness. 

${\cal Z}_b$ is obtained from ${\cal Z}_{r, \, g}$ by taking the 
limit, $\Delta = \Delta^* \to 0$; 
\begin{equation} 
\ln {\cal Z}_b = \frac{1}{2} \ln {\cal Z}_{r, \, g} \, 
\rule[-3mm]{.14mm}{8.5mm} 
\raisebox{-2.85mm}{\scriptsize{$\; \Delta = \Delta^* = 0$}} \, . 
\label{blue1} 
\end{equation}
Then we concentrate on evaluating ${\cal Z}_{r, \, g}$; 
\begin{equation} 
\ln {\cal Z}_{r, \, g} = \frac{1}{2} \ln \mbox{Det} \left( 
- \beta G^{- 1} \right) \, , 
\label{kihon} 
\end{equation}
the computation of which is given in Appendix D. 
\subsubsection*{Thermodynamic potential}
From Eqs. (\ref{zeroth}), (\ref{arya}), (\ref{finn}) (cf. Eq. 
(\ref{finn4})), and Eq. (\ref{blue1}), we obtain for the 
thermodynamic potential, 
\begin{eqnarray}
\Omega &=& - \frac{T}{V} \, \ln {\cal Z} \nonumber \\ 
& = & \frac{\sigma^2}{4 G_S} + \frac{|\Delta|^2}{4 G_D} - N_f 
\int \frac{d^{\, 3} p}{(2 \pi)^3} \sum_{\rho, \, \sigma = \pm} 
\left[ E_{p, \, \rho}^{(\sigma)} + 2 T \ln \left( 1 + e^{- \beta 
E_{p, \, \rho}^{(\sigma)}} \right) \right. \nonumber \\ 
&& \left. + \frac{1}{2} \left\{ E_{p, \, \rho}^{(\sigma)} + 2 T \ln 
\left( 1 + e^{- \beta E_{p, \, \rho}^{(\sigma)}} \right) 
\right\}_{\Delta = 0} \right] \, , 
\label{finn2}
\end{eqnarray}
where $N_f = 2$ is the number of flavor-degrees of freedom, and 
\begin{equation}
E_{p, \, \rho}^{(\sigma)} = \left[ 
\left(E_p + \sigma \frac{\mu_+ + \mu_-}{2} \right)^2 + |\Delta|^2 
\right]^{1 / 2} + \rho \frac{\mu_+ - \mu_-}{2} \, . 
\label{finn3} 
\end{equation}
The three momentum integral in Eq. (\ref{finn2}) is ultraviolet 
divergent, and, as usual, we introduce a momentum cut-off parameter 
$\Lambda$.   
The zero-temperature limit ($T \to 0$) of $\Omega$ reads 
\begin{equation}
\Omega (T = 0) = \frac{\sigma^2}{4 G_S} + 
\frac{|\Delta|^2}{4 G_D} - N_f \int \frac{d^{\, 3} p}{(2 \pi)^3} 
\sum_{\rho, \, \sigma = \pm} \left[ 
\, \rule[-1.5mm]{.14mm}{5.5mm} E_{p, \, \rho}^{(\sigma)} 
\, \rule[-1.5mm]{.14mm}{5.5mm} 
+ \frac{1}{2} 
\, \rule[-1.5mm]{.14mm}{5.5mm} 
E_{p, \, \rho}^{(\sigma)} 
\, \rule[-1.5mm]{.14mm}{5.5mm} 
\raisebox{-2.mm}{\scriptsize{$\; \Delta = 0$}} 
\right] \, . 
\label{tdp0}
\end{equation}

$\Omega$ in Eqs. (\ref{finn2}) and (\ref{tdp0}) yields the gap equations: 
\begin{equation} 
\frac{\partial \Omega}{\partial m} = 0 \, , \;\;\;\;\;\;\; 
\frac{\partial \Omega}{\partial \Delta} = 0 \;\;\;\;\;\;\;\;\; 
(\Delta \equiv |\Delta|) \, . 
\label{gap} 
\end{equation} 
A solution ($m$, $\Delta$) to Eq. (\ref{gap}) is a local minimum 
point of $\Omega$, when the following conditions are met: 
\begin{eqnarray}
&& \frac{\partial^2 \Omega}{\partial m^2} \frac{\partial^2 
\Omega}{\partial |\Delta^2|} - \left( \frac{\partial^2 
\Omega}{\partial m \partial \Delta} \right)^2 > 0 \, , \nonumber \\ 
&& \frac{\partial^2 \Omega}{\partial m^2} 
+ \frac{\partial^2 \Omega}{\partial |\Delta|^2} > 0 \, . 
\label{kensa}
\end{eqnarray}
\subsubsection*{Polarized quark-number density}
Let $n^{(\pm)}$ be the \lq\lq positive (negative) magnetic moment'' 
net quark-number density. $n^{(\pm)}$ is computed from $\Omega$ through 
partial differentiation with respect to $\mu_\pm$: 
\begin{eqnarray}
n^{(\pm)} (\mu_+, \mu_-, T) & = & - \frac{\partial \Omega}{\partial 
\mu_\pm} \;\;\; \left( \equiv n^{(\pm)}_{r, \, g} + n^{(\pm)}_b 
\right) \, , 
\label{qnumber} \\ 
n^{(\pm)}_{r, \, g} (\mu_+, \mu_-, T) & = & N_f \int \frac{d^{\, 3} 
p}{(2 \pi)^3} \theta (\Lambda - p) \sum_{\sigma = \pm} \left[ 
\frac{\sigma E_{p, \, \sigma}}{\sqrt{E_{p, \, \sigma}^2 + 
|\Delta|^2}} \right. \nonumber \\ 
& & \left. - \sum_{\rho = \pm} \left( \frac{\sigma E_{p, \, 
\sigma}}{\sqrt{E_{p, \, \sigma}^2 + |\Delta|^2}} \pm \rho \right) 
n_F \left( E_{p, \, \rho}^{(\sigma)} \right) \right] \, , 
\label{qnumber1} 
\\ 
n^{(\pm)}_b (\mu_+, \mu_-, T) & = & N_f \int \frac{d^{\, 3} 
p}{(2 \pi)^3} \theta (\Lambda - p) \left[ n_F (E_p - \mu_\pm)-  n_F 
(E_p + \mu_\pm) \right] \, , 
\label{qnumber2} 
\end{eqnarray}
where $E_{p, \, \sigma} \equiv E_p + \sigma \bar{\mu}$ ($\bar{\mu} 
\equiv (\mu_+ + \mu_-)/2$) and $n_F (x) = 1 / (e^{\beta x} + 1)$ is 
the Fermi distribution function. In the limit $T \to 0$, $n^{(\pm)}$ 
turns out to 
\begin{eqnarray}
n^{(\pm)}_{r, \, g} (T = 0) &=& N_f \int \frac{d^{\, 
3} p}{(2 \pi)^3} \theta (\Lambda - p) \nonumber \\ 
&& \times \left[ \frac{\bar{\mu} + E_p}{\sqrt{(\bar{\mu} + E_p)^2 + 
|\Delta|^2}} + \frac{\bar{\mu} - E_p}{\sqrt{(\bar{\mu} - E_p)^2 + 
|\Delta|^2}} \right. \nonumber \\ 
&& - \theta \left( \frac{(\delta \mu)^2}{4} - |\Delta|^2 \right) \, 
\theta \left( E_p - \bar{\mu} + \sqrt{\frac{(\delta \mu)^2}{4} - 
|\Delta |^2} \right) \nonumber \\ 
&& \left. \times \left( \frac{\bar{\mu} - E_p}{\sqrt{(\bar{\mu} - 
E_p)^2 + | \Delta |^2}} \mp 1 \right) \theta \left( \bar{\mu} + 
\sqrt{\frac{(\delta \mu)^2}{4} - |\Delta |^2} - E_p \right) \right] 
\, , \\ 
n^{(\pm)}_b (T = 0) & = & N_f \int \frac{d^{\, 3} 
p}{(2 \pi)^3} \theta (\Lambda - p) \theta (\mu_\pm - E_p) \, , 
\label{qnumber0} 
\end{eqnarray}
where $\delta \mu \equiv \mu_+ - \mu_-$. 
\subsection{Spin-polarized quark matter}
The counterpart to Eq. (\ref{kakuyo}) is 
\begin{equation} 
\left( G_0^\pm \right)^{- 1} = \sum_{\tau, \, s = \pm} \left( p_0 + 
\tau E_p \pm \mu_s \right) \gamma_0 {\bf P}_s^{(\tau)} 
(- \vec{p}) \, . 
\label{kakuyo1} 
\end{equation}
$\Omega$ is computed in a similar manner as in Appendix D and we 
obtain the same form as above, i.e., Eq. (\ref{finn2}) with Eq. 
(\ref{finn3}). 
\setcounter{section}{3}
\setcounter{equation}{0}
\def\theequation{\mbox{\arabic{section}.\arabic{equation}}} 
\section{Numerical analysis} 
In this section, through numerical analyses, we will analyze the 
phase structure of the unpolaried and polarized quark matters at 
zero and finite temperatures. How the chiral condensate, diquark 
condensate, and the degree of polarization compete with one another 
will be discussed. 

We first fix the parameters of the model as in \cite{mei}: The 
current quark mass $m_0 = 5.5$ MeV, the three-momentum cut-off 
$\Lambda = 0.637$ GeV, and the coupling constant $G_S = 5.32$ 
GeV$^{- 2}$, which reproduces the properties of pion in vacuum. As 
mentioned above, although the standard value of $G_D / G_S$ is $3 / 
4$, we leave it as a free parameter. 
\subsection{Unpolarized quark matter} 
\subsubsection{$T = 0$}

\begin{figure}[h]
\begin{center}
\begin{tabular}{cc}
\includegraphics[width=6.7cm,clip]{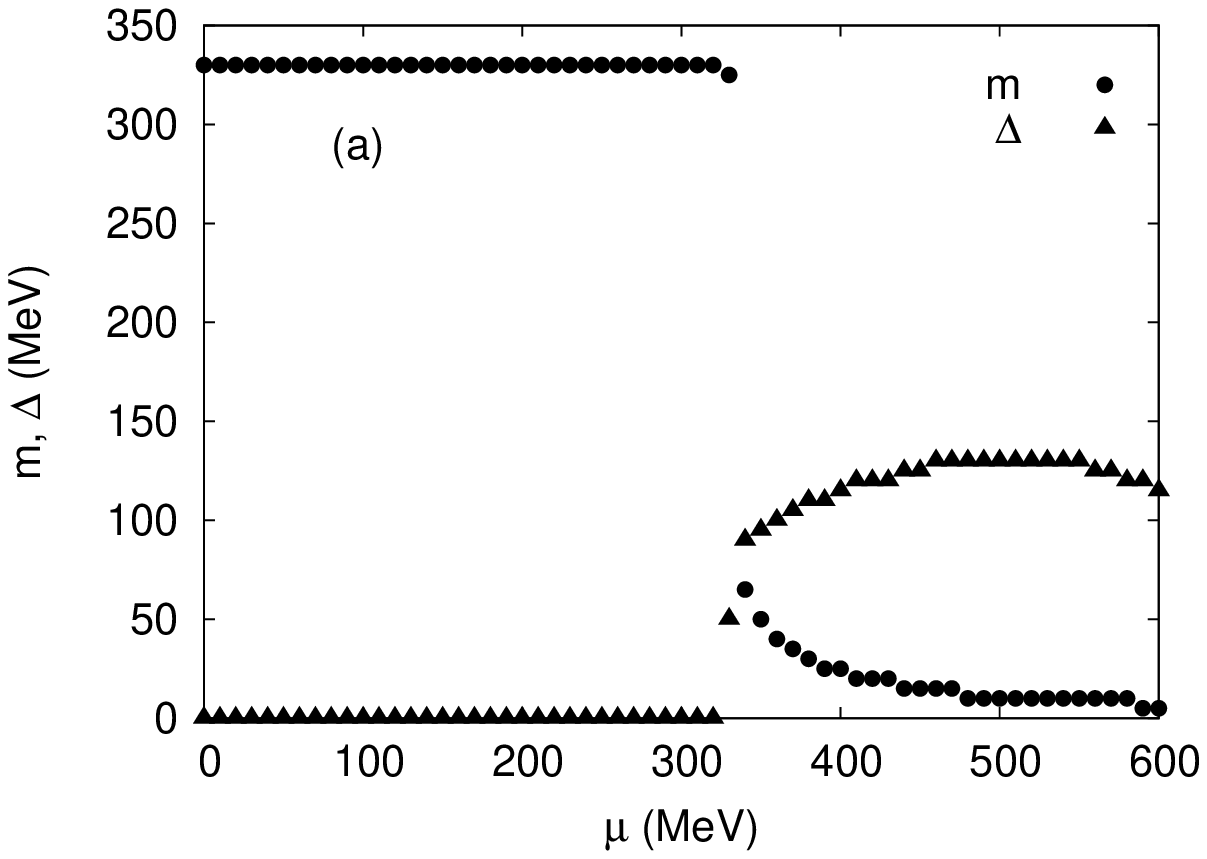} &
\includegraphics[width=6.7cm,clip]{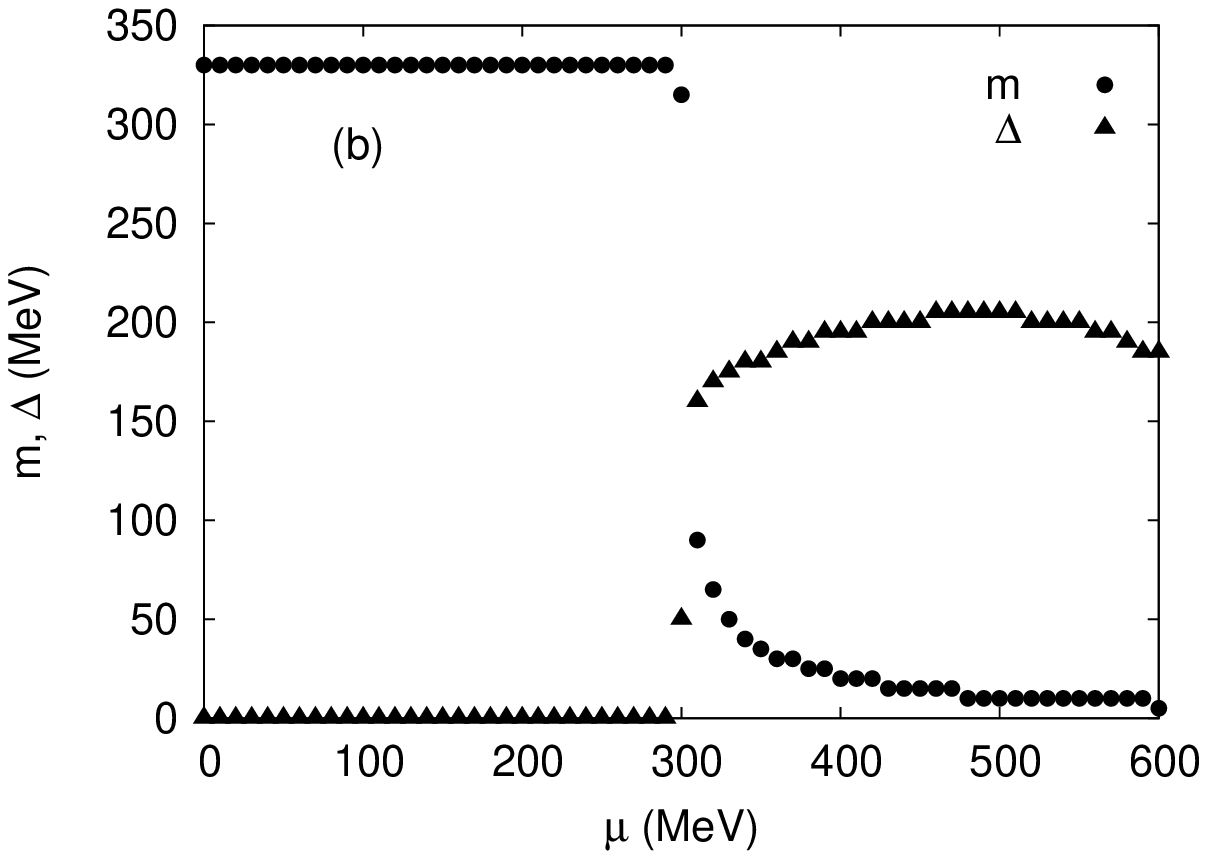} \\
\includegraphics[width=6.7cm,clip]{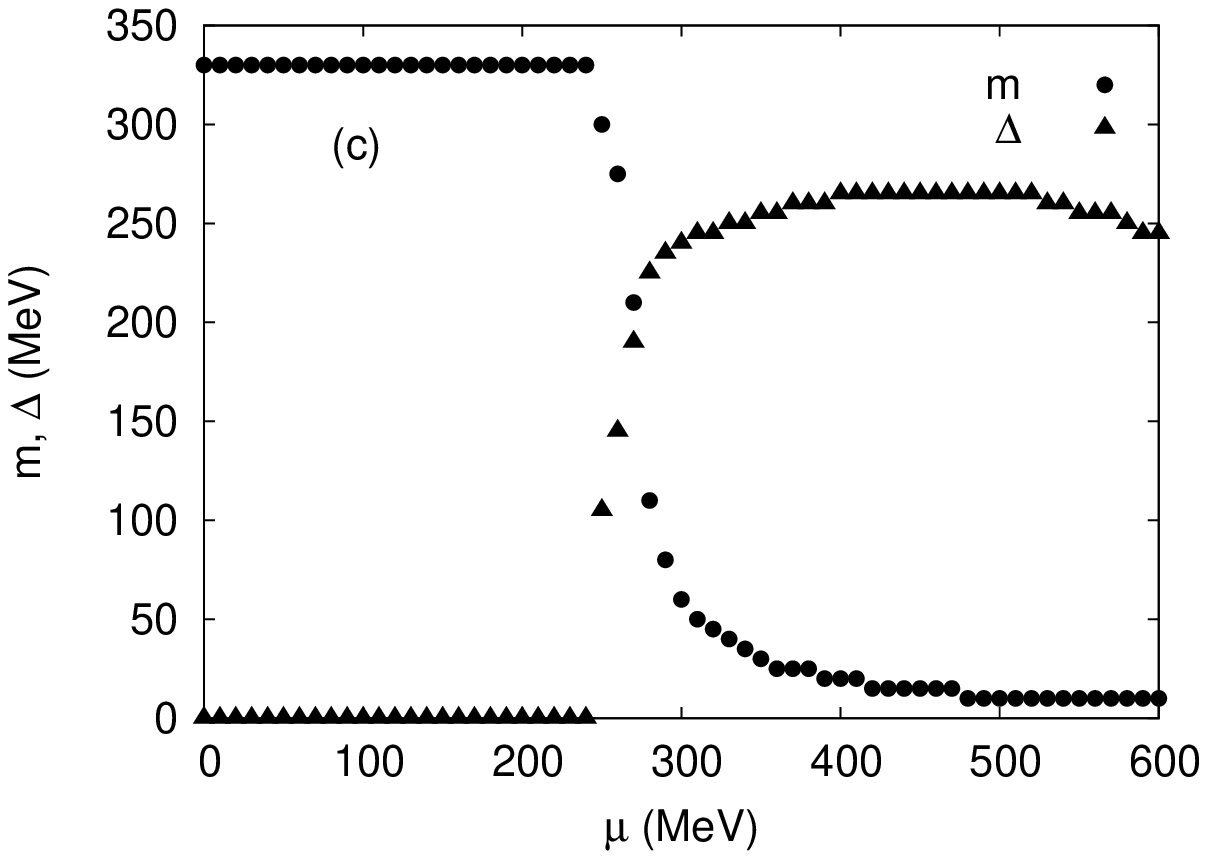} &
\includegraphics[width=6.7cm,clip]{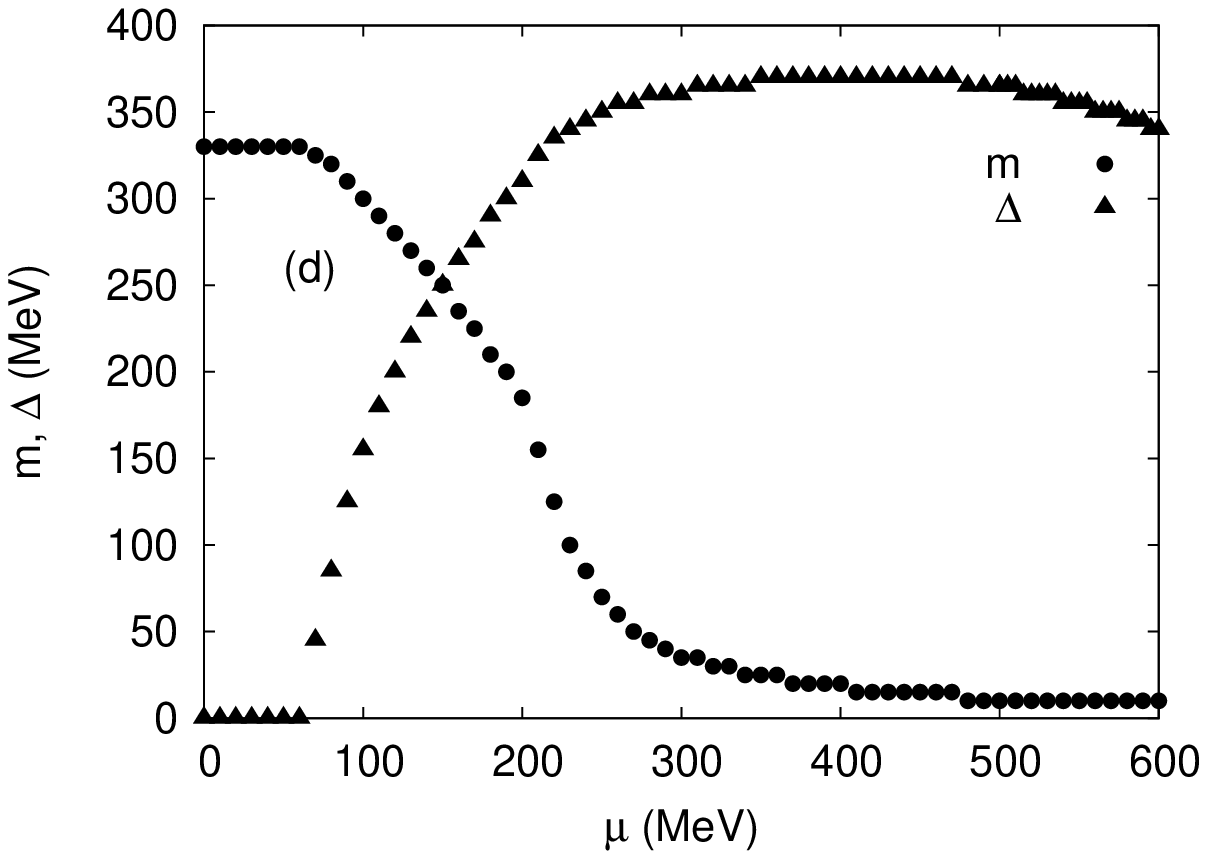} \\
\end{tabular}
 \caption{Plots of m and $\Delta$ as functions of the chemical potential $\mu$ for $T=\delta \mu = 0$.}
 {(a), (b), (c), and (d) correspond to $G_{D}/G_{s}=$ 3/4, 1.0, 1.2, and 1.5, respectively.}
\end{center}
\end{figure}

Throughout in the sequel of this section, values of all dimensionful 
quantities are given in unit of MeV. In \cite{mei}, the two gaps $m$ 
and $\Delta (\equiv |\Delta|)$ at $T = 0$ are displayed for various 
values of $G_D / G_S$. For completeness, we reproduce them in Fig. 
1. The panel (a) corresponds to the standard value $G_D / G_S = 3 / 4$, 
while (b), (c) and (d) correspond, as in \cite{mei}, to $G_D / G_S = 
1.0, 1.2,$ and $1.5$, respectively. Close observation on Fig. 1 is 
given in \cite {mei}, which we briefly recapitulate here. 

Fig. 1 (a) ($G_D / G_S = 3 / 4$) shows that the chiral phase 
transition and the color superconducting (diquark) phase 
transition take place at nearly the same chemical potential 
$\mu_\chi \simeq \mu_\Delta \simeq 330$. These two phase 
transitions are of first order. There remains a small chiral 
condensate in the color superconducting (CSC) phase $\mu > 
\mu_\Delta$, which is a relic of the explicit chiral symmtery 
breaking ($m_0 \neq 0$). 

We define $\mu_\Delta$, as usual, to be the value of the chemical 
potential at which the diquark condensate starts to appear. We 
define $\mu_\chi$ to be the chemical potential at which $m$ starts 
to decrease from nearly the constant value (rather than the chemical 
potential at which the chiral symmtery is restored). From Fig. 1 (b) 
- (d), we see that $\mu_\Delta \simeq \mu_\chi$ and, as $G_D / G_S$ 
inceases, $\mu_\Delta \simeq \mu_\chi$ decreases. The chiral 
condensate in the coexistence phase is due to the dynamical symmtery 
breaking. 

Here we like to add one finding. Within the accuracy of numerical 
computation, the value of $\mu$ at which the quark-number density, 
$n$, vanishes coincides with $\mu_\Delta$, i.e., $n (\mu, T = 0) = 
0$ for $\mu \leq \mu_\Delta$. 
\subsubsection{$T \neq 0$}
For the shake of copmarison to the case of $\delta \mu \neq 0$ in 
the next subsection, we display here the results for $T \neq 0$ and 
$\delta \mu = 0$. 

\begin{figure}[h]
\begin{center}
\begin{tabular}{cc}
\includegraphics[width=6.7cm,clip]{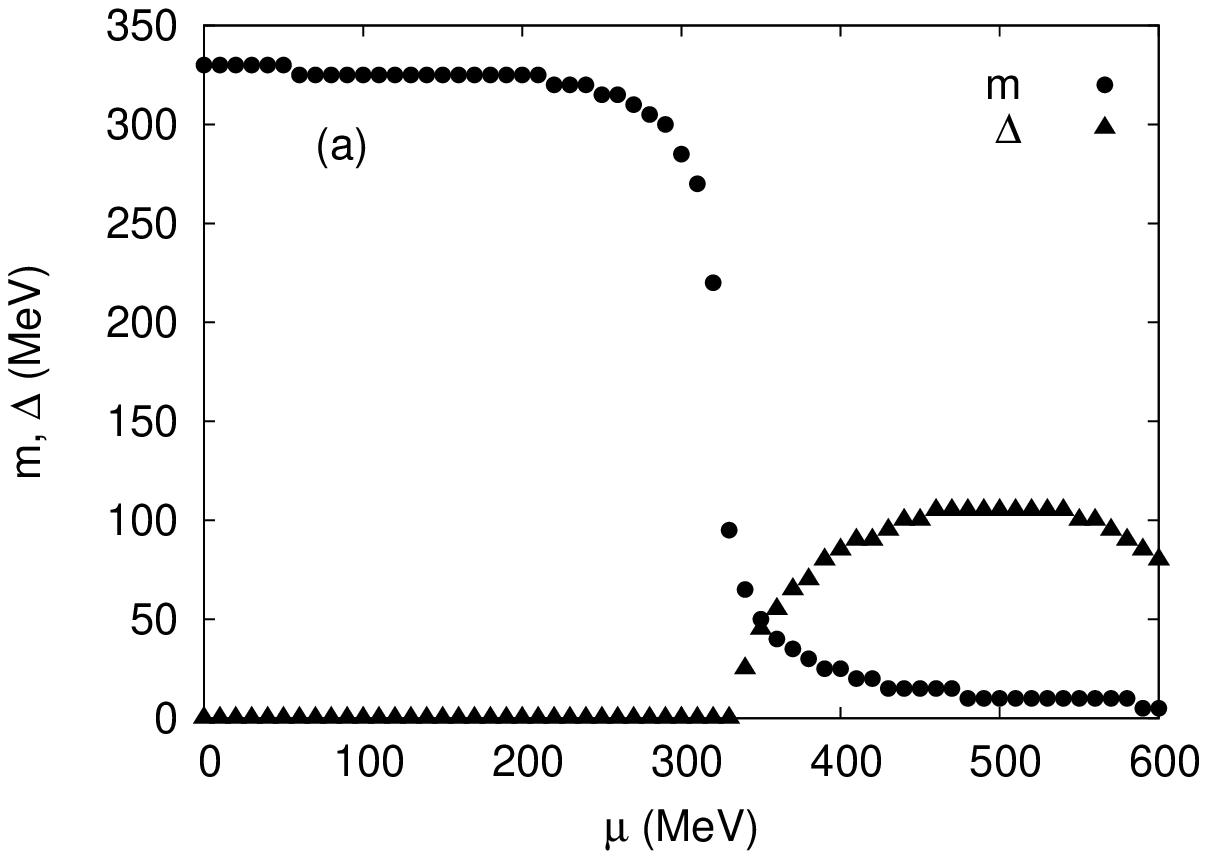} &
\includegraphics[width=6.7cm,clip]{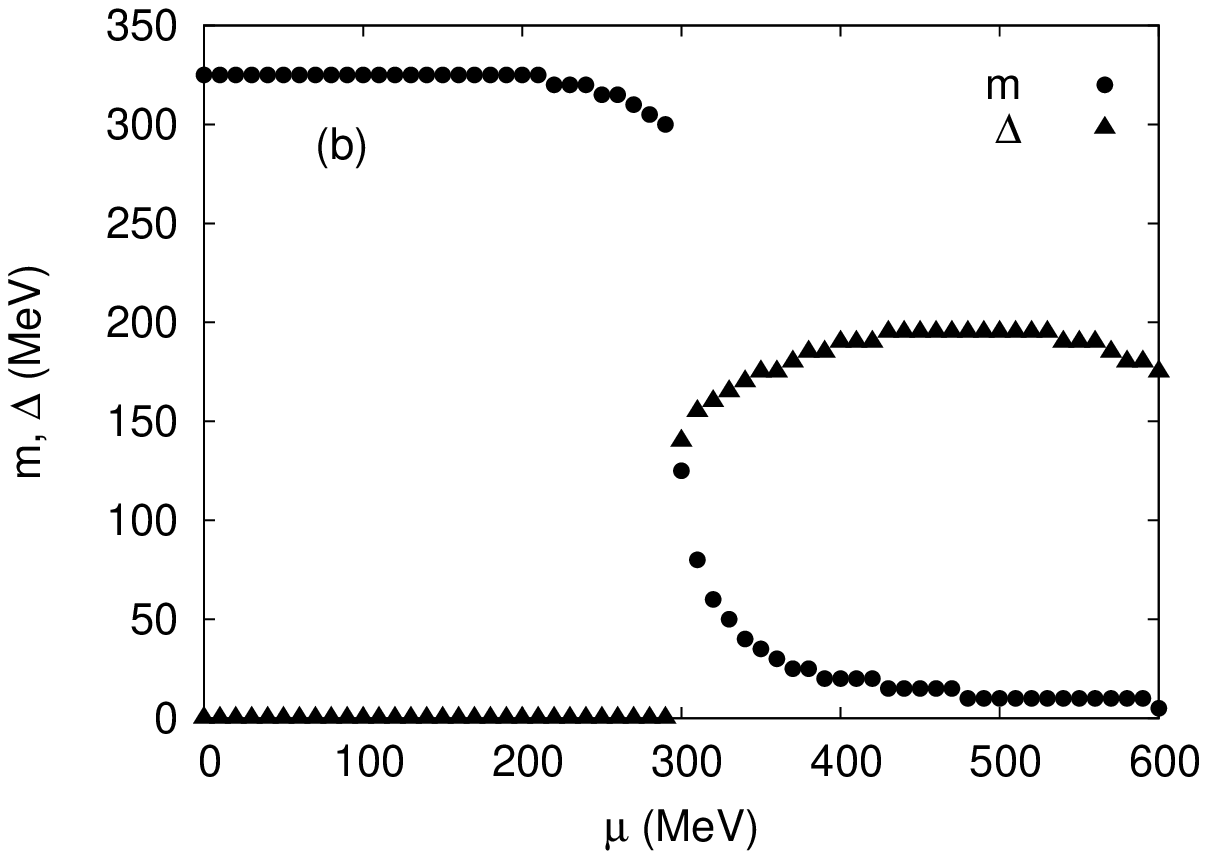} \\
\includegraphics[width=6.7cm,clip]{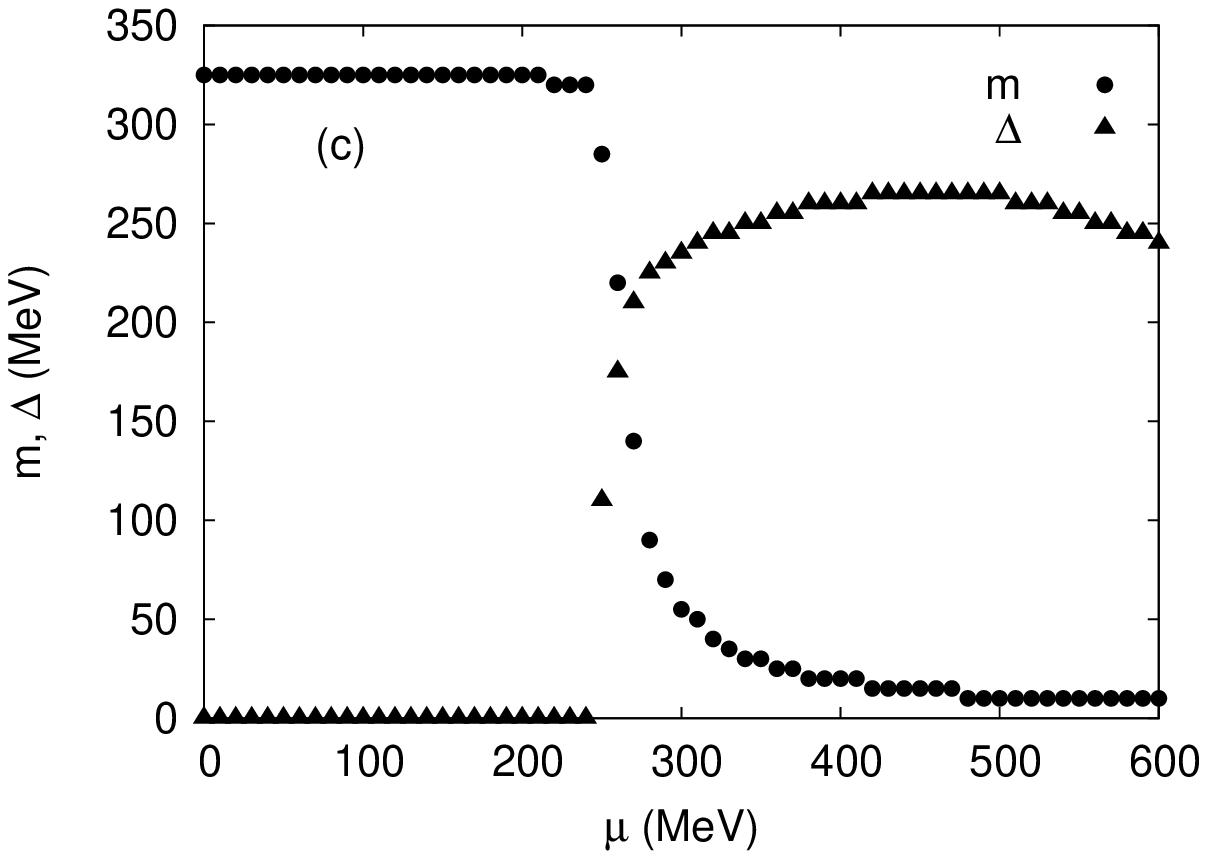} &
\includegraphics[width=6.7cm,clip]{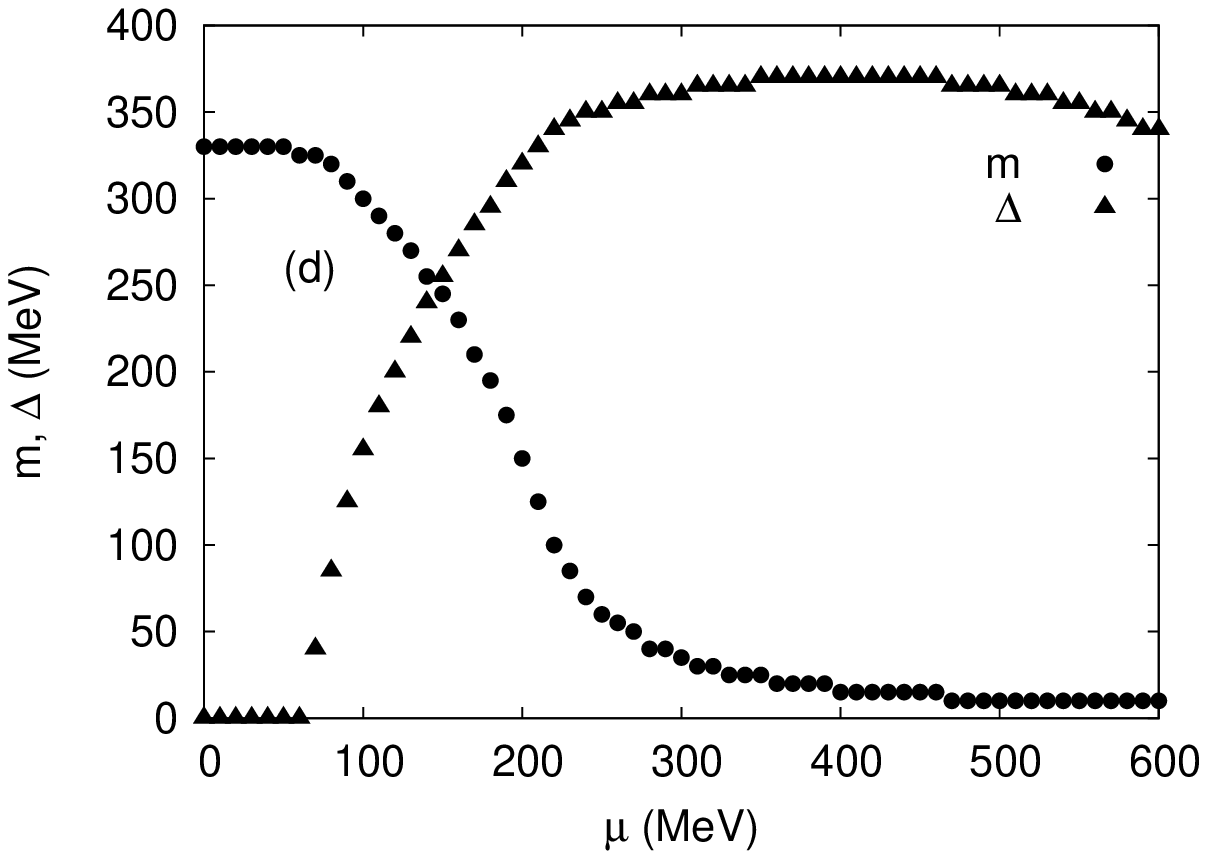} \\
\end{tabular}
\caption{ The same as Fig. 1 for $T=50$.}
\end{center}
\end{figure}

In Fig. 2, we plot $m$ and $\Delta$ at $T = 50$ for the same values 
of $G_D / G_S$ as in Fig. 1. Fig. 2 (a) shows that the value of 
$\mu_\Delta$ is nearly the same as in the $T = 0$ case, and 
$\mu_\chi < \mu_\Delta$. Both phase transitions are of second order. 
$\Delta$ in the region $\mu > \mu_\Delta$ is smaller than that in 
the $T = 0$ case. Comparing Fig. 1 (b) - (d) with Fig. 2 (b) - (d), 
we can make similar observation, besides that $\Delta$'s in the 
region $\mu > \mu_\Delta$ at $T = 50$ are not appreciably smaller 
than their $T = 0$ counterparts. 

In the following we refer the region where $\Delta = 0$ ($m = O 
(m_0)$) to as the hadronic-phase (CSC-phase) region, 
while the region in between them \textit{simply} to as the double-broken 
phase. 

\begin{figure}[h]
\begin{center}
\begin{tabular}{cc}
\includegraphics[width=6.7cm,clip]{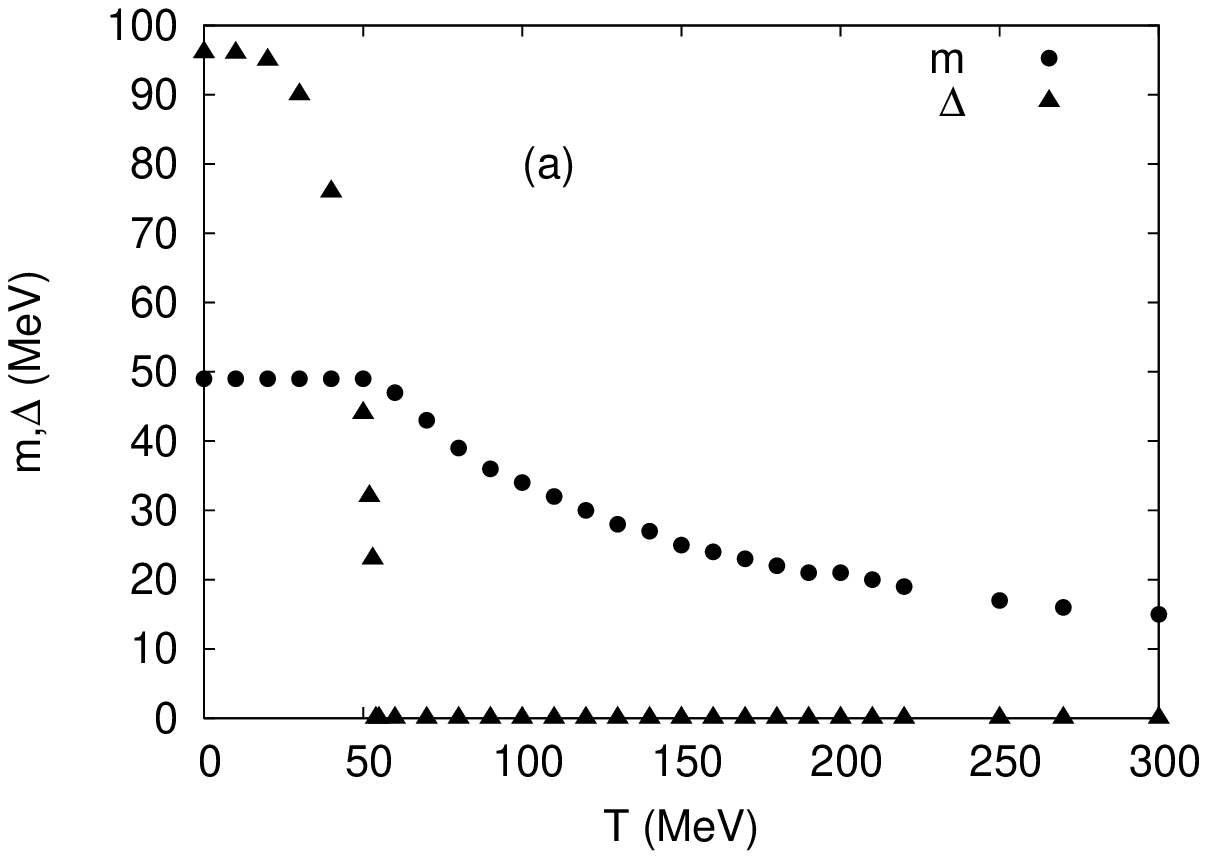} &
\includegraphics[width=6.7cm,clip]{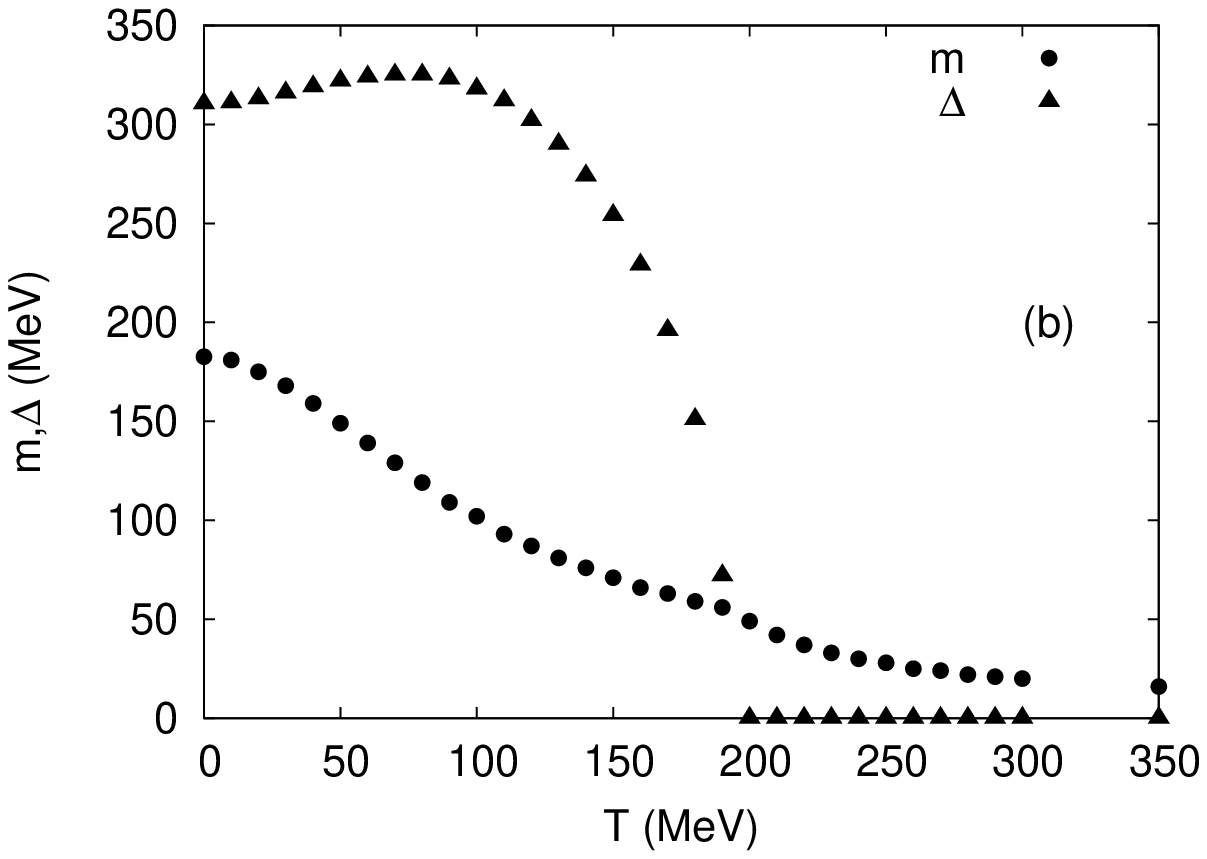} \\
\end{tabular}
\caption{$m, \Delta$ against T for (a) $(G_{D}/G_{S}, \mu )=(3/4, 350)$ and (b) $(G_{D}/G_{S}, \mu )=(1.5, 200)$.}
\end{center}
\end{figure}

Let us see the $T$ dependence of $m$ and $\Delta$. In Fig. 3, 
choosing the double-broken phase region, we plot $m$ and $\Delta$ 
against $T$. The panel (a) [(b)] corresponds to $G_D / G_S = 3/4$ 
[$1.5$] and $\mu = 350$ $[200]$. We see that $\Delta$ decreases as 
$T$ increases and becomes zero at the critical temperature 
$T_\Delta$. Above $T_\Delta$, the color normal phase is realized. 
With respect to the chiral symmetry, the panel (a) shows that, as 
$T$ increases, $m$ starts to decrease at $T \simeq T_\Delta$, and 
the chiral symmetry is restored gradually. On the other hand, in the 
case of panel (b), $m$ starts to decrease from the beginning ($T = 
0$) and the chiral symmetry is restored gradually. We have observed 
the similar behavious for other values of $\mu$. 

Finally, in Fig. 4, we display the phase diagrams. The panels (a), 
(b), and (c) correpond to $G_D/ G_S = 3 / 4$, $1.2$, and $1.5$, 
respectively. 

\begin{figure}[h]
\begin{center}
\begin{tabular}{ccc}
\includegraphics[width=6cm,clip]{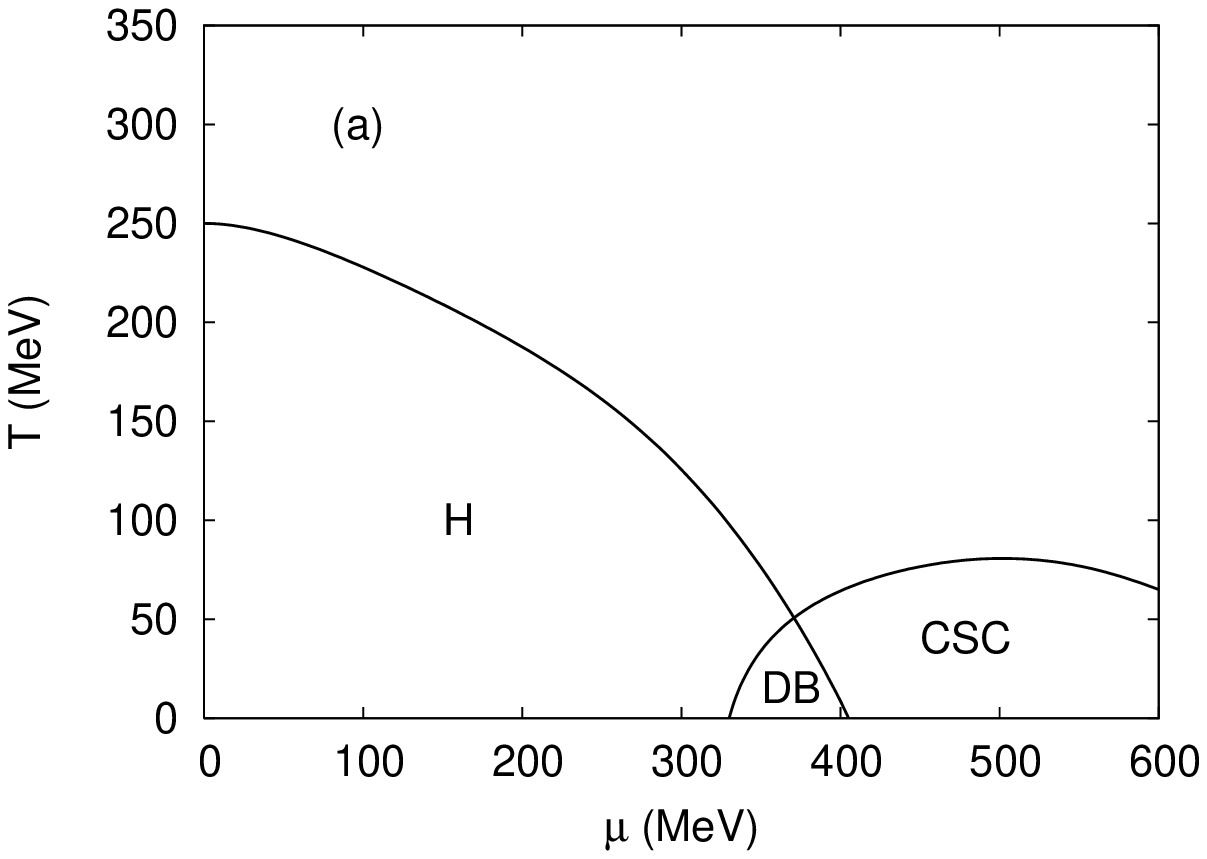} &
\includegraphics[width=6cm,clip]{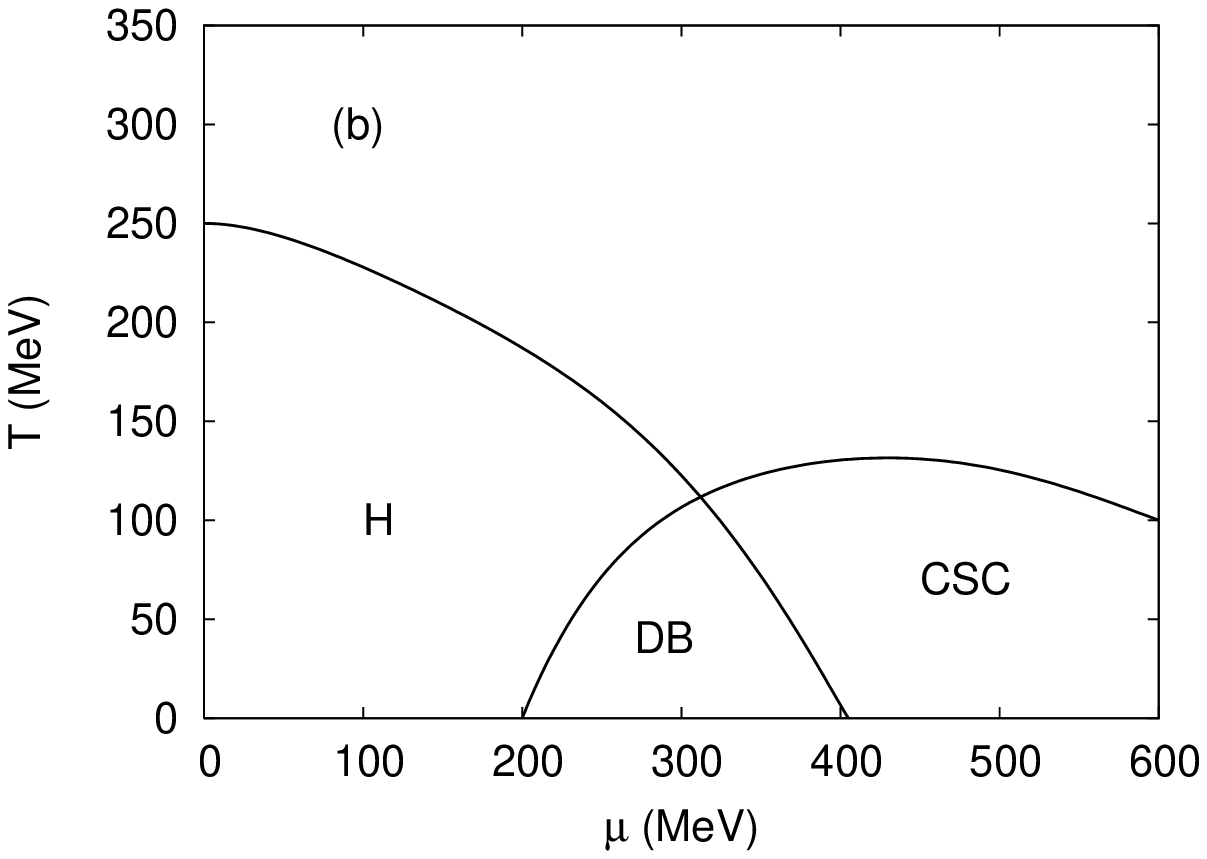} &
\includegraphics[width=6cm,clip]{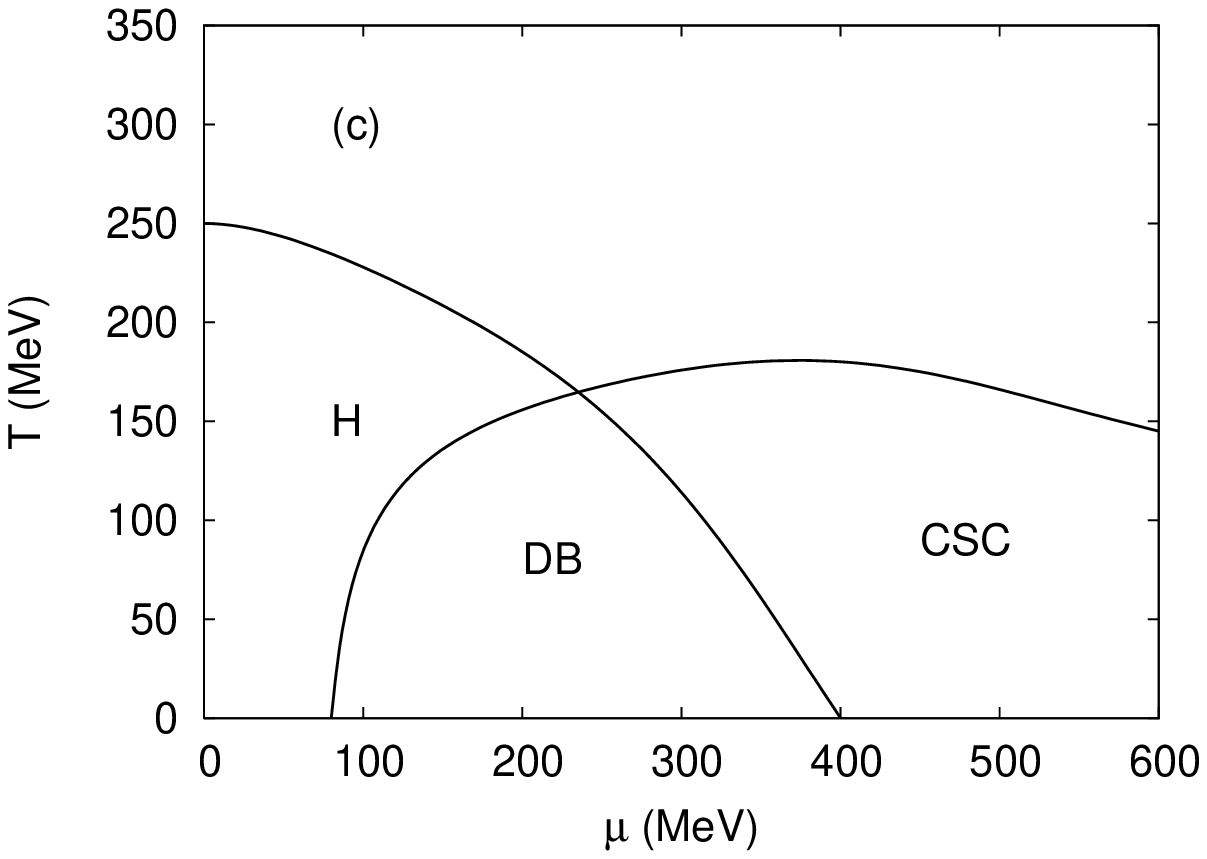} \\
\end{tabular}
 \caption{Phase diagrams for $G_D/ G_S = 3 / 4$ (a), 1.2 (b), and 1.5 (c).} 
 {H, CSC, and DB stand for hadronic, color-super-conducting, and double-broken phases, respectively.}
\end{center}
\end{figure}

\subsection{Polarized quark matter} 
We have, among others, three parameters $\mu_+$, $\mu_-$, and $T$. 
We difine $\mu$ through 
\begin{equation}
n_b^{(+)} (\mu_+, \mu_-, T) \, \rule[-3mm]{.14mm}{8.5mm} 
\raisebox{-2.85mm}{\scriptsize{$\; \mu_+ = \mu_- = \mu$}} = 
\frac{1}{6} \, \left[ n^{(+)} (\mu_+, \mu_-, T) + n^{(-)} 
(\mu_+, \mu_-, T) \right] \, , 
\label{maa}
\end{equation}
where $n^{(\pm)} (\mu_+, \mu_-, T)$ is as in Eqs. (\ref{qnumber}) - 
(\ref{qnumber2}). The procedure of numerical computation for various 
quantities goes as follows. 
\begin{description}
\item{1)} We first fix $T$, $\mu (\geq 0)$, and $\mu_- (\geq 0)$. 
\item{2)} Then, we solve two gap equatios, Eq. (\ref{gap}), and Eq. 
(\ref{maa}) simultaneously, and obtain the solution for ($\mu_+, m, 
\Delta$), and compute $\delta \mu = \mu_+ - \mu_-$. \item{3)} 
Finally, we check if the two conditions (\ref{kensa}) are met. 
\end{description}
\subsubsection{$\delta \mu$ dependence of $m$ and $\Delta$} 
{\em Double-broken phase}: We first study the parameter region where 
the unpolarized quark matter is in the double-broken phase. In Fig. 
5, we plot $m$ and $\Delta$ agaisnt $\delta \mu$ for different 
values of ($G_D / G_S$, $T$, $\mu$): The panels (a), (b), (c), and (d) 
correspond to $(3 / 4, 0, 400)$, $(3 / 4, 50, 400)$, $(1.5, 0, 
300)$, and $(1.5, 50, 300)$, respectively. Fig. 5 (a) shows that, as 
$\delta \mu$ increases, the first-order phase transition occurs at 
$\delta \mu = \delta \mu_p \simeq 220$. For $\delta \mu > \delta 
\mu_p$, the system is in the color normal phase. At $\delta \mu = 
\delta \mu_p$, a gap appears in $m$. Fig. 5 (b) shows similar 
behaviors as Fig. 5 (a) , but no gap appears in $m$ at $\delta \mu = 
\delta \mu_p$. Figs. 5 (c) and (d) shows that, in the case of $G_D / 
G_S = 1.5$, both $m$ and $\Delta$ are nearly independent of $\delta 
\mu$, and no phase transition occurs. We have carried out the same 
computation for different parameter values and found qualitatively 
the same results. 

\begin{figure}[h]
\begin{center}
\begin{tabular}{cc}
\includegraphics[width=6.5cm,clip]{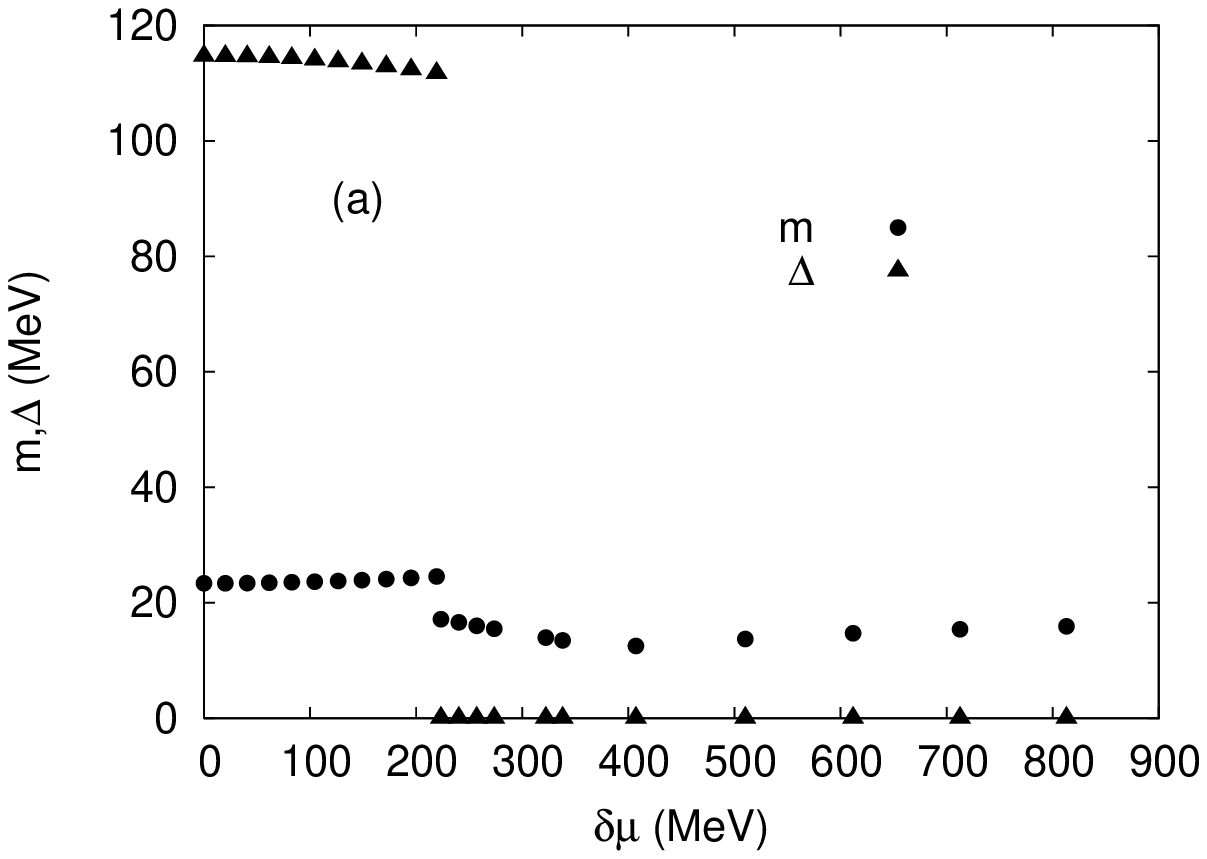} &
\includegraphics[width=6.5cm,clip]{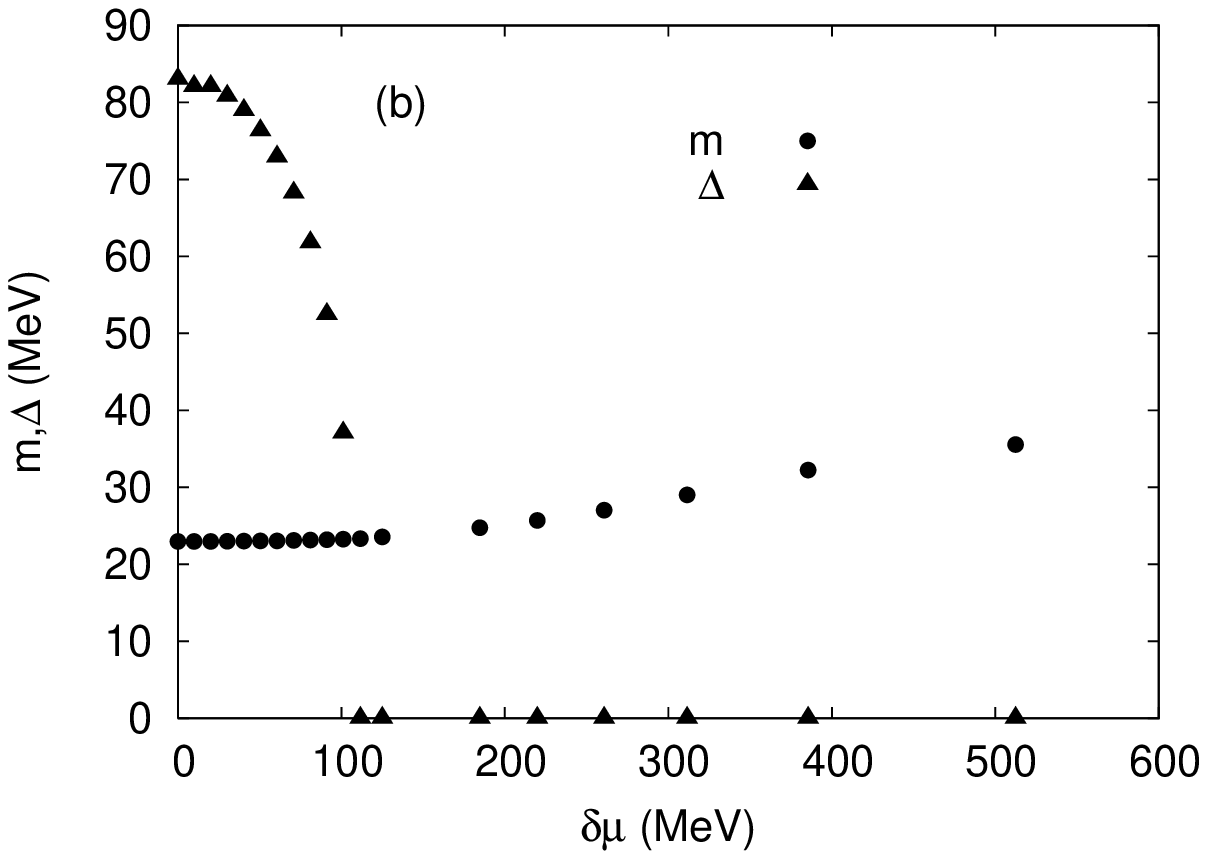} \\
\includegraphics[width=6.5cm,clip]{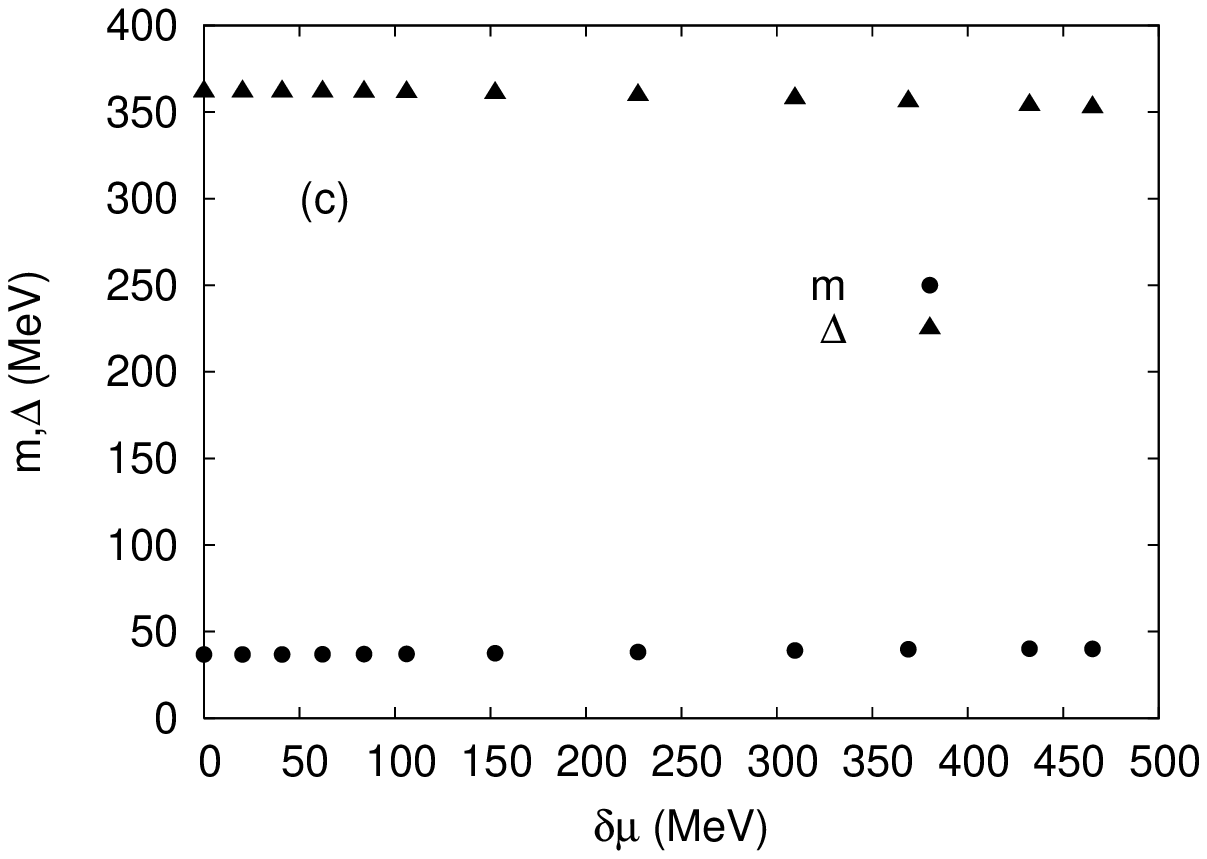} &
\includegraphics[width=6.5cm,clip]{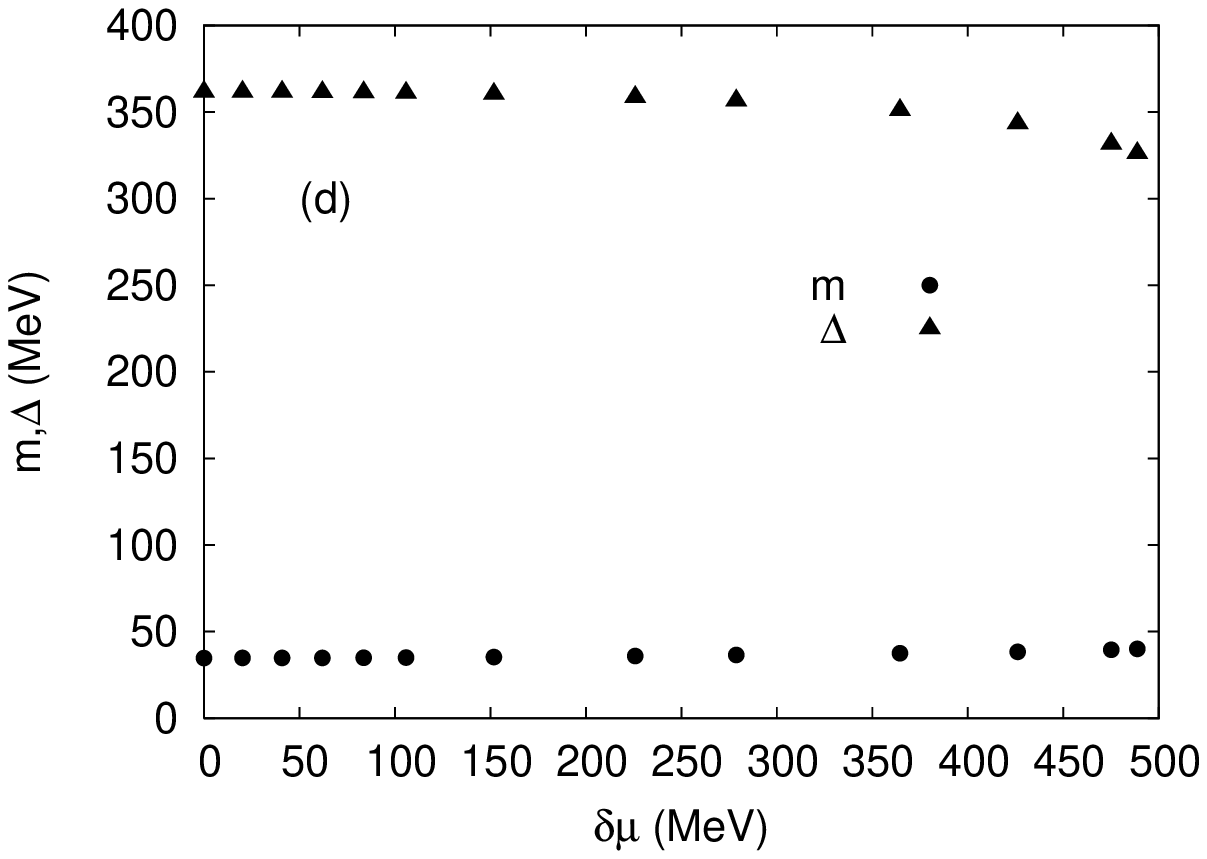} \\
\end{tabular}
\caption{m and $\Delta$ against $\delta\mu$. (a), (b), (c), and (d) correspond to 
($G_D / G_S$, T, $\mu$)=(3/4, 0, 400),} 
{(3/4, 50, 400), (1.5, 0, 300), and (1.5, 50, 300), respectively.}
\end{center}
\end{figure}

\begin{figure}[h]
\begin{center}
\begin{minipage}{.45\linewidth}
  \includegraphics[width=7cm,clip]{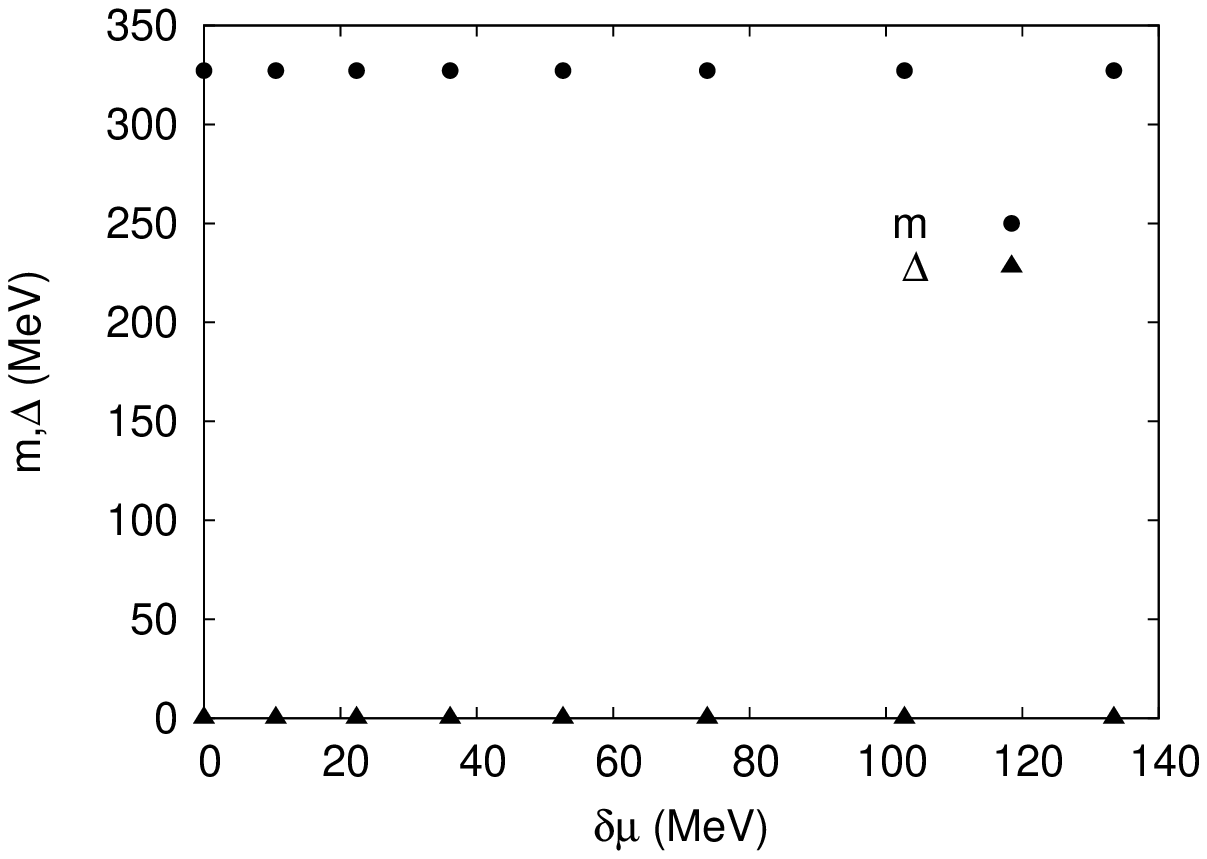}
  \caption{m and $\Delta$ against $\delta\mu$ for ($G_D / G_S$, T, $\mu$)}{=(1.0, 50, 100).}
  \end{minipage}
  \hspace{2.0pc}
  \begin{minipage}{.45\linewidth}
  \includegraphics[width=7cm,clip]{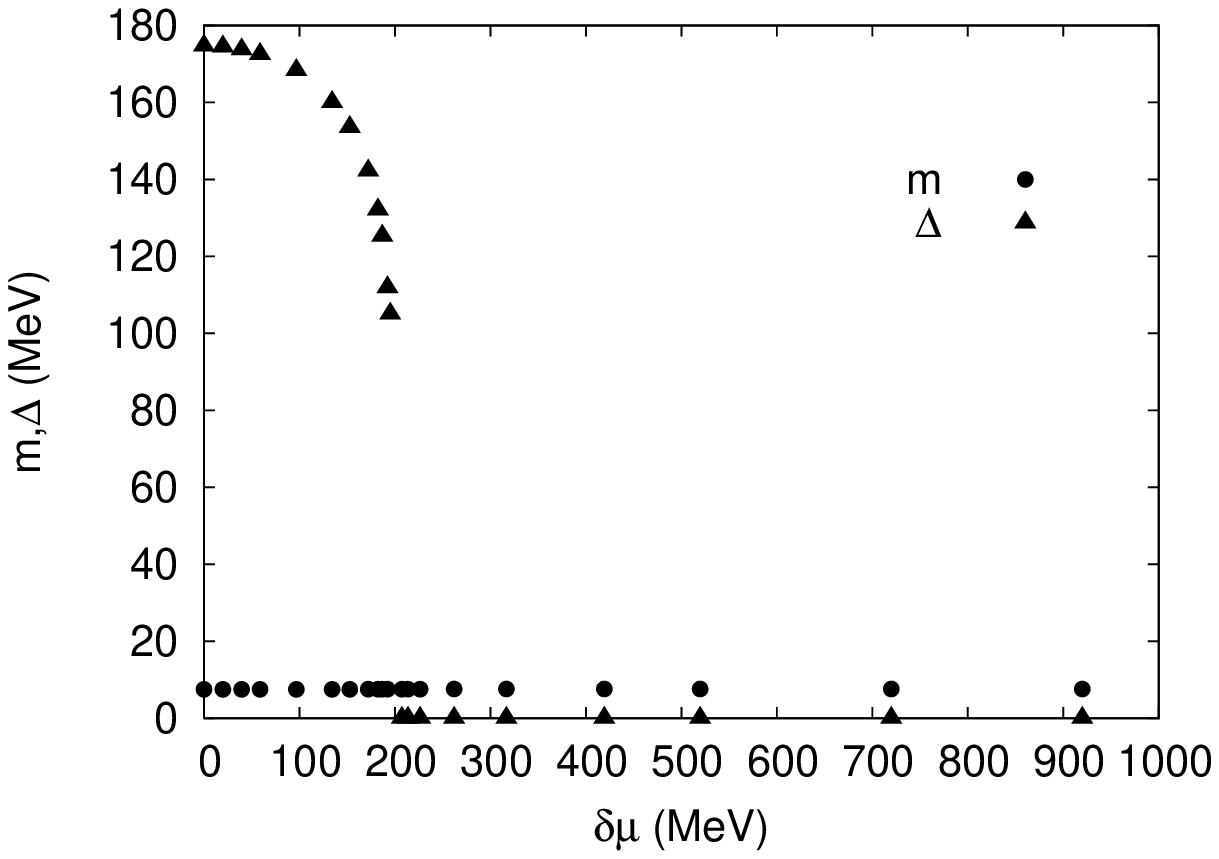}
  \caption{m and $\Delta$ against $\delta\mu$ for ($G_D / G_S$, T, $\mu$)}{=(1.0, 50, 600).} 
\end{minipage}
\end{center}
\end{figure}

{\em Hadronic phase}: We study the parameter region where 
the unpolarized quark matter is in the hadronic phase. In 
Fig. 6, $m$ and $\Delta$ are plotted against $\delta \mu$ for 
$(G_D/ G_S, T, \mu) = (1.0, 50, 100)$ (cf. Fig. 2 (b)). We see that 
$m$ and $\Delta$ $(= 0)$ are nearly independent of $\delta \mu$ and 
the quark matters are in the color normal phase for all $\delta 
\mu$. We have carried out the same computation choosing different 
parameter values and found qualitatively the same results. 

{\em CSC phase}: Finally, we study the parameter region where 
the unpolarized quark matter is in the CSC phase. In Fig.7, we plot 
$m$ and $\Delta$ against $\delta \mu$ for $(G_D/ G_S, T, \mu) = (1.0, 
50, 600)$ (cf. Fig. 2 (b)). As $\delta \mu$ increases, a phase 
transition occurs at $\delta \mu = \delta \mu_p$ and the quark 
matter turns out to the color normal phase, while (small) $m$ is 
nearly independent of $\delta \mu$. We have carried out the same 
computation choosing different parameter values and found 
qualitatively the same results. 
\subsubsection{Thermodynamic potential} 
Here we compute the thermodynamic potential $\Omega (\mu, T; \delta 
\mu)$ as a function of $\delta \mu$, for different values of $(\mu, 
T)$, and discuss their physical implications. 

Two typical behaviors of $\Omega$ are shown in Fig. 8. In the 
system with $\Omega$ as in the panel (a), the \lq\lq polarized'' or 
\lq\lq ferromagnetic'' phase is realized, and we refer such an 
$\Omega$ to as the \lq\lq ferromagnetic (FM)'' type. On the other 
hand, in the system with $\Omega$ as in the panel (b), the unpolarized 
phase is realized, and we refer such an $\Omega$ to as the \lq\lq 
normal (N)'' type. 

\begin{figure}[h]
\begin{center}
\begin{tabular}{cc}
\includegraphics[width=6.7cm,clip]{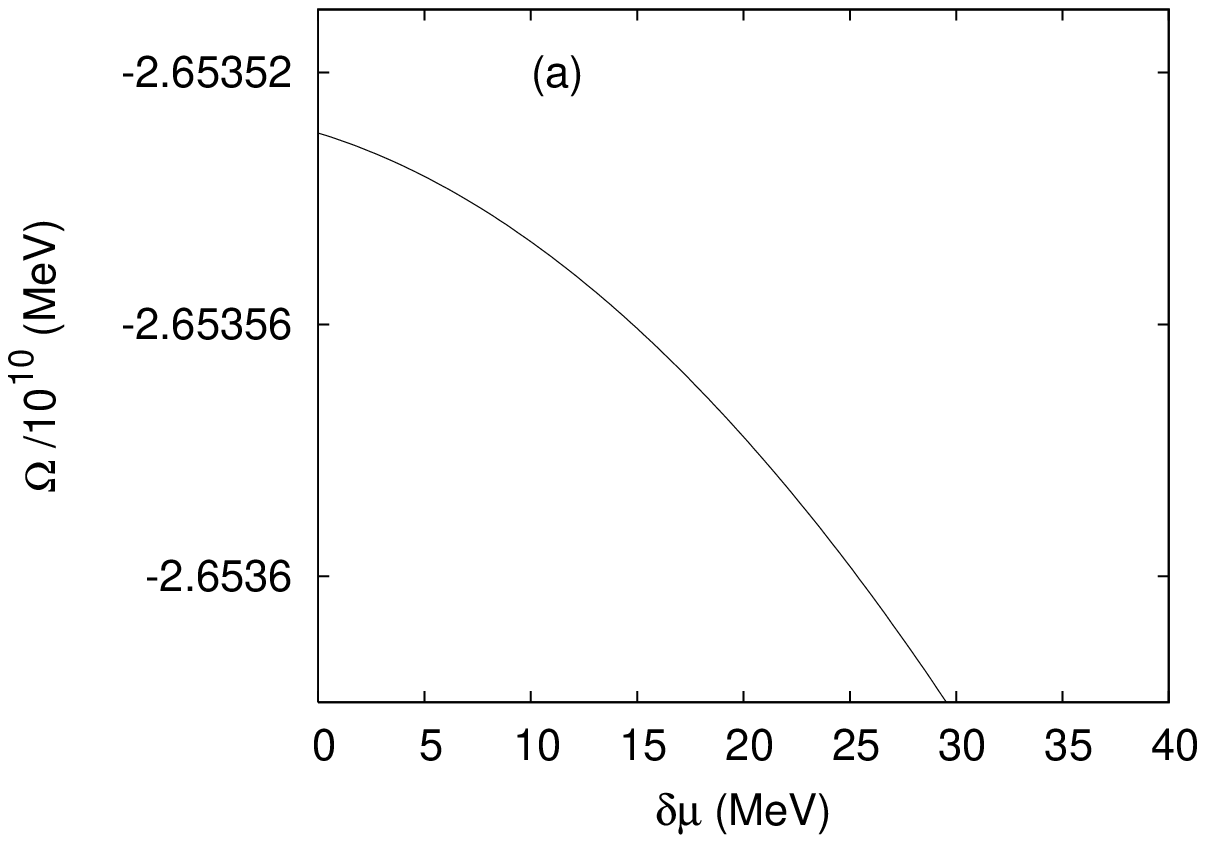} &
\includegraphics[width=6.7cm,clip]{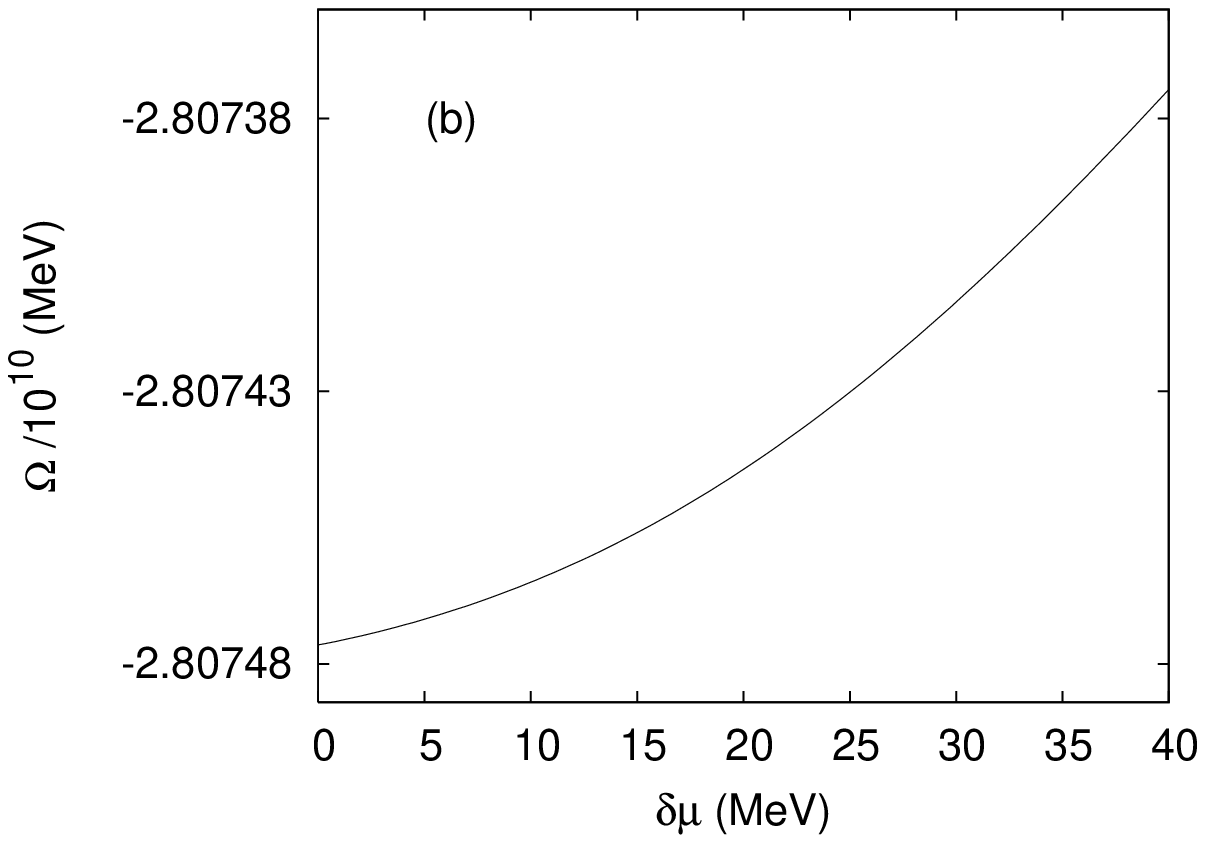} \\
\end{tabular}
\caption{$\Omega$(T=0) against $\delta \mu$ for $\mu =400$. The panels (a) and (b) correspond to 
$G_{D}/G_{S}=3/4$ and 1.2, respectively.}
\end{center}
\end{figure}

\begin{figure}[h]
\begin{center}
\begin{tabular}{cc}
\includegraphics[width=6.7cm,clip]{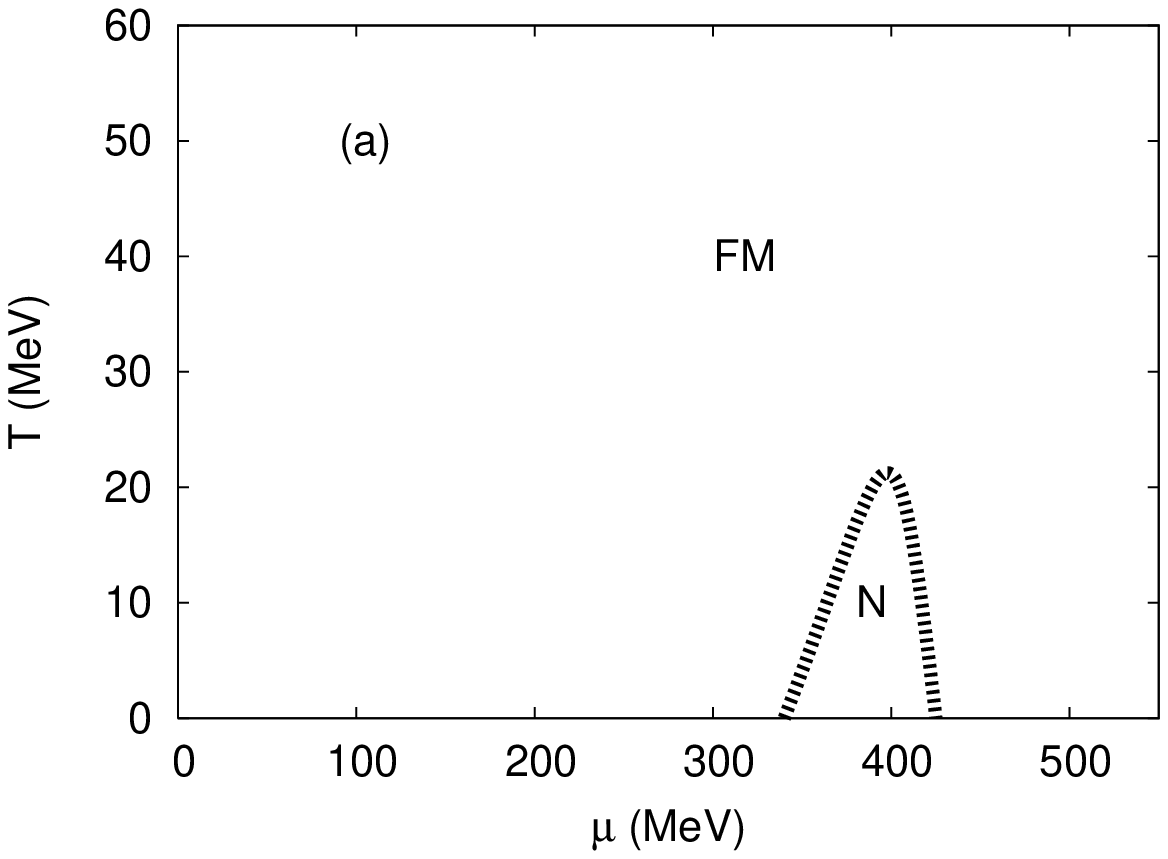} &
\includegraphics[width=6.7cm,clip]{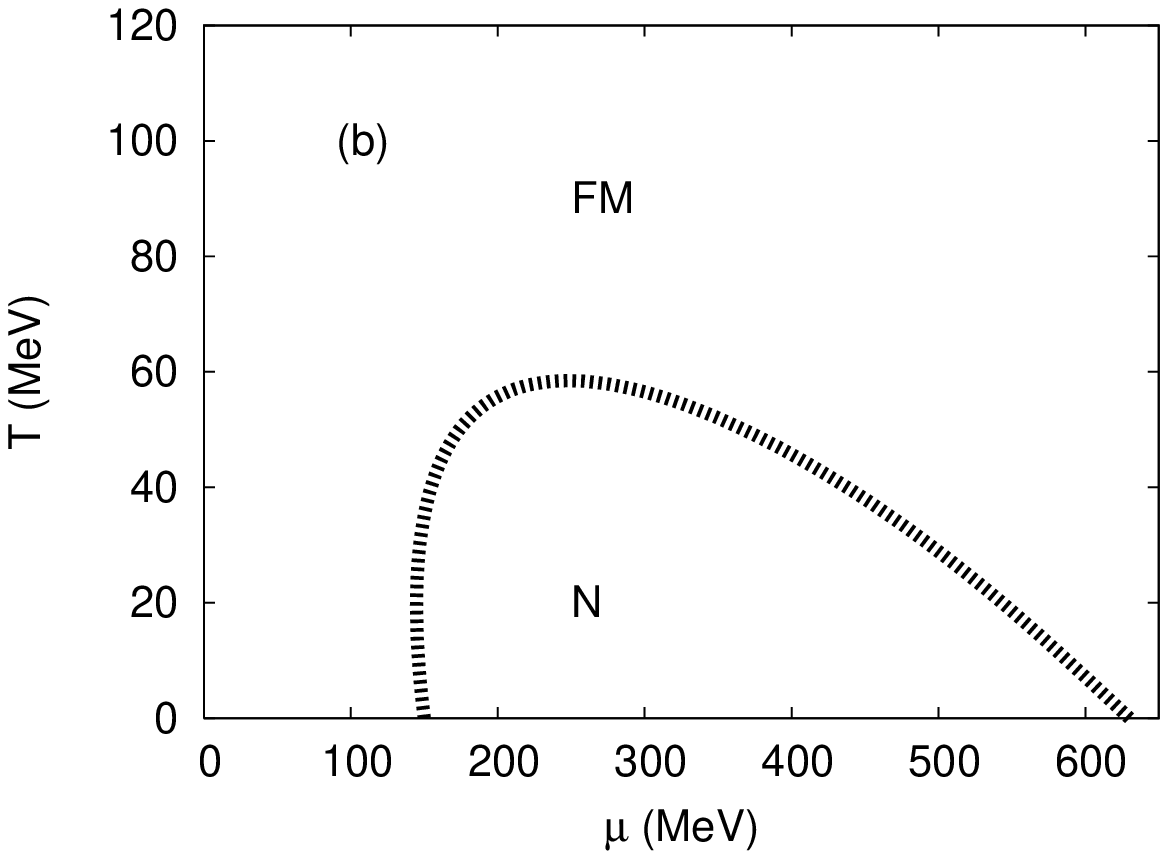} \\
\end{tabular}
\caption{Phase diagrams with respect to polarization. The panels (a) and (b) correspond to 
$G_{D}/G_{S}=1.2$ and 1.5, respectively.}{ FM and N stand for the $\lq\lq$ferromagnetic" and $\lq\lq$normal" phase, respectively.}
\end{center}
\end{figure}

Through numerical computation of $\Omega$, we have found the 
following results. 
\begin{description}
\item{1)} For $G_D / G_S \lesssim 1.15$, $\Omega$ is of the FM-type 
in whole region of $(\mu, T)$. As an illustraion, we depict in Fig. 
8 (a) $\Omega (\mu = 400, T = 0; \delta \mu)$ as a function of 
$\delta \mu$ for $G_D / G_S = 3 / 4$. 
\item{2)} For $G_D / G_S \gtrsim 1.15$, there appears a window in 
the $(\mu, T)$-plane, in which $\Omega$ is of the N-type. We show in 
Fig. 9 the phase diagrams in the $(\mu, T)$-plane. The panel (a) 
[(b)] corresponds to $G_D / G_S = 1.2$ $[1.5]$. As an illustration, 
we plot in Fig. 8 (b) $\Omega (\mu = 400, T = 0; \delta \mu)$ 
against $\delta \mu$ for 
$G_D / G_S = 1.2$, which is of the N-type. 
\end{description}

To see how $\Omega$ changes through the transition region from 
the FM-type region to the N-type region, we display $\Omega(\mu = 
330, T = 0; \delta \mu)$ and $\Omega(\mu = 340, T = 0; \delta \mu)$ 
for $G_D / G_S = 1.2$ in Figs. 10 (a) and (b), respectively. Fig. 10 
(a) indicates that, for $(G_D / G_S, \mu, T) = (1.2, 330, 0)$, a 
metastable state appears at $\delta \mu = 0$, while Fig. 10 (b) 
shows that, for $(G_D / G_S, \mu, T) = (1.2, 340, 0)$, a stable 
state appears at intermediate $\delta \mu$. 

\begin{figure}[h]
\begin{center}
\begin{tabular}{cc}
\includegraphics[width=6.7cm,clip]{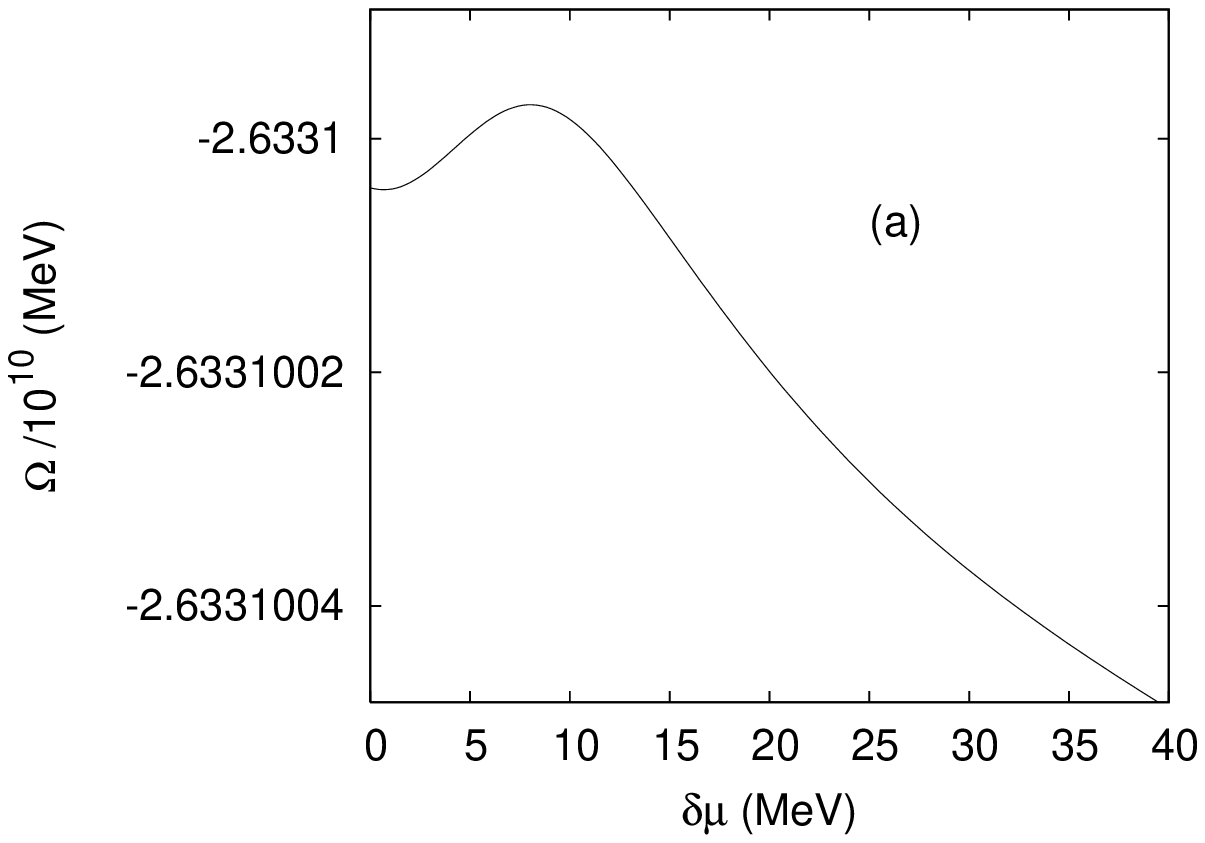} &
\includegraphics[width=6.7cm,clip]{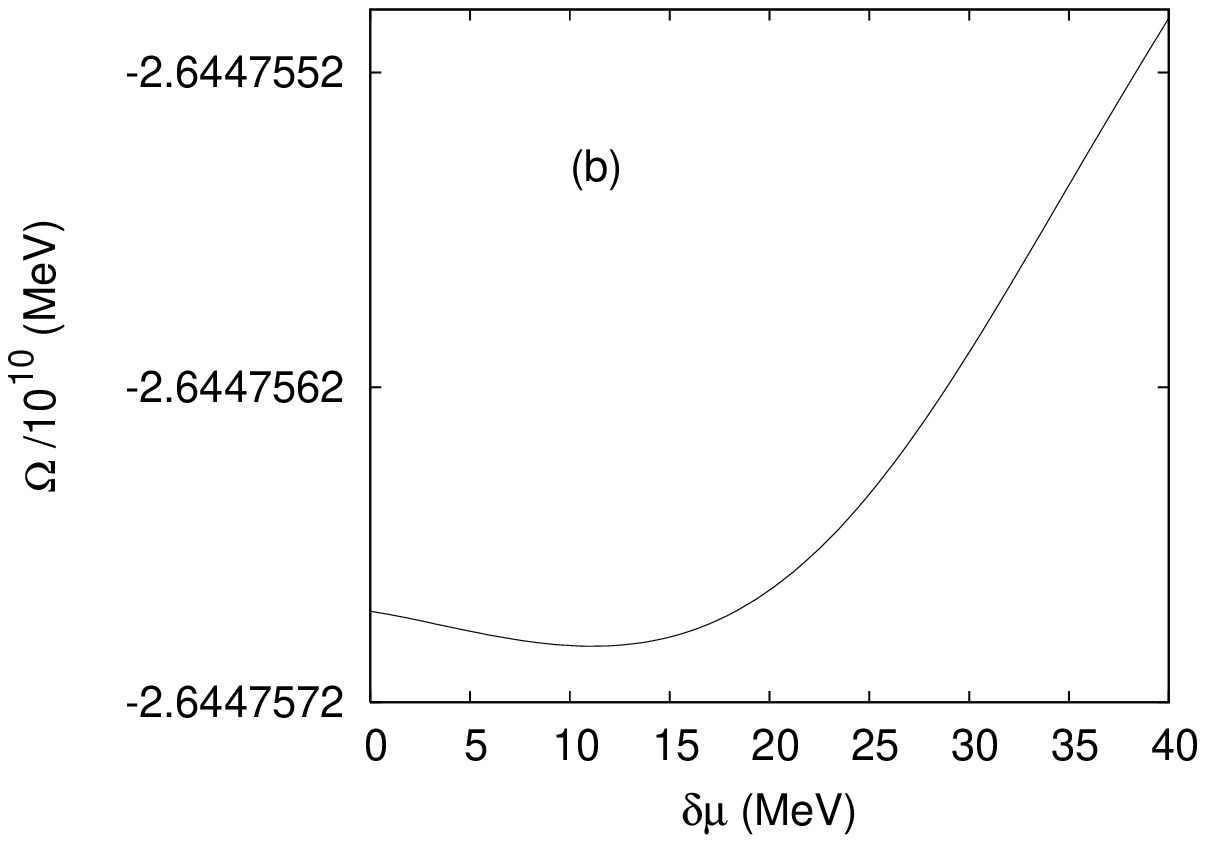} \\
\end{tabular}
\caption{$\Omega$(T=0) against $\delta \mu$ for $G_{D}/G_{S}=1.2$. The panels (a) and (b) correspond to 
$\mu =330$ and 340, respectively.}
\end{center}
\end{figure}

Numerical analyses show that, in the region where $\Delta = 0$, 
$\Omega$ decreases as $\delta \mu$ increases, so that $\Omega$ is 
of the FM-type. Since $\Delta = 0$ at high temperature (cf. Figs. 3 and 4), 
$\Omega$ is of the FM-type at high $T$. For the normal phase to 
appear, $\Delta \neq 0$ is necessary. 
\setcounter{section}{4}
\def\theequation{\mbox{\arabic{section}.\arabic{equation}}} 
\setcounter{equation}{1}
\section{Summary and discussion} 
In this paper, generalizing the analysis in \cite{tatsumi}, we have 
dealt with unpolarized and polarized two-flavor quark matters at 
zero and finite temperatures. 

We have used the extended Nambu--Jona-Lassinio model and employed 
the mean-field approximation. The model contains two coupling 
constants, the quark-antiquark coupling constant $G_S$ and the 
diquark coupling constant $G_D$. Although $G_S$ and $G_D$ have the 
relation $G_D = (3 / 4) G_S$, we have treated, as in \cite{mei}, 
$G_D / G_S$ as a free parameter. 

The unpolarized case at $T = 0$ are fully analyzed in \cite{mei}. 
In addition to it, we have analyzed the unpolarized case at finite 
temperatures through computing the thermodynamic potential. The 
result is summarized as the phase diagrams in the $(\mu, T)$-plane 
in Fig. 4. The structure of the diagrams are qualitatively the same 
for all values of $G_D/ G_S$ considered in this paper. 

We have dealt with two types of polarized two-flavor quark matters, 
i.e., the \lq\lq magnetic-moment''-polarized and the \lq\lq 
spin''-polarized quark matters. For describing such systems, the 
extended Nambu--Jona-Lassionio model is generalized to treat the 
chiral condensate, diquark condensate, and the degree of 
polarization on an equal footing. We have shown that, within the 
mean-field approximation, the form of the thermodynamic potential 
$\Omega$ is the same for both the \lq\lq magnetic-moment''-polarized 
and the \lq\lq spin''-polarized cases. 

We have found that, as in the cases of other quantities, the $\delta 
\mu$ dependence of $\Omega$ heavily depends on the value of $G_D / 
G_S$. For small $G_D / G_S \lesssim 1.15$, $\Omega$ is of the \lq\lq 
ferromagnetic''-type, while, for large $G_D / G_S \gtrsim 1.15$, 
there appears a window in the $(\mu, T)$-plane, where $\Omega$ is of 
the \lq\lq normal''-type. We have found interesting behaviors of $\Omega$ 
in the transition regions between the \lq\lq ferromagnetic"-type and the 
\lq\lq normal"-type regions. We have seen that, at high temerature and/or density, 
the \lq\lq ferromagnetic'' phase is energetically favored over the \lq\lq normal" phase, 
which is in conflict with intuition. This fact may suggests that the 
extended Nambu--Jona-Lasinio model supplemented with the mean-field 
approximation is not applicable to the polarized quark matter in 
such regions. 

The analysis presented in this paper is the \lq\lq first-stage 
analysis''. Closer investigation is to be done in various 
directions. Among others, are 1) inclusion of the residual 
interactions, and 2) incorporation of electron, positron, and gluons 
into the NJL model. Also interesting is to apply the method 
in this paper to quark matters in different phases, 
such as the gapless charge-neutral 2SC phase, the color-flavor-loced 
phase, the non-uniformly condensed phases \cite{sadz}, and so on. 
\begin{appendix}
\setcounter{equation}{1}
\setcounter{section}{0}
\def\theequation{\mbox{\Alph{section}.\arabic{equation}}}
\section{The role of the projectors ${\cal P}_s^{\tau}$ ($\tau = 
\pm, s = \pm$)}
We start with briefly reviewing the spin content of $q (x)$. We 
employ the box normalization by confining the system within a cube 
and introduce the periodic boundary condition to make the 
single-particle plane-wave basis. $q (x)$ in the interaction picture 
is expanded, in standard notations, as 
\[ 
q (x) = \sum_{\bf p} \sum_{s = \pm} \sqrt{\frac{m_0}{VE}} \left[ 
a_s (\vec{p}) u_s (\vec{p}) e^{- i p \cdot x} + 
b_s^\dagger (\vec{p}) v_s (\vec{p}) e^{i p \cdot x} \right] \, , 
\] 
where ${\cal P}_s (\vec{p}) u_{s'} (\vec{p}) = \delta_{s s'} u_s 
(\vec{p})$ and ${\cal P}_s (\vec{p}) v_{s'} (\vec{p}) = 
\delta_{s s'} v_s (\vec{p})$. The third component of the spin 
operator reads $S^z = \frac{1}{2} \int d^3 x \, q^\dagger (x) 
\sigma^3 q (x)$. Expectation values of $S^z$ in one quark- and one 
antiquark- states are computed, in respective order, as (cf. 
\cite{tatsumi}) 
\begin{eqnarray}
\langle 0 | a_\pm (\vec{p}) S^z a_\pm^\dagger (\vec{p}) | 0 \rangle 
&=& \frac{m_0}{2 E_p} u^\dagger (\vec{p}) {\cal P}_\pm^\dagger 
(\vec{p}) \sigma^3 {\cal P}_\pm (\vec{p}) u (\vec{p}) \nonumber \\ 
&=& \frac{1}{2} \mbox{Tr} \left( {\bf P}_\pm^{(-)} (- \vec{p}) 
\sigma^3 \right) \nonumber \\ 
&=& \pm \frac{m_0}{2 E_p} n^3 (\vec{p}) \, , 
\label{tasu1} \\ 
\langle 0 | b_\pm (\vec{p}) S^z a_\pm^\dagger (\vec{p}) | 0 \rangle 
&=& - \frac{m_0}{2 E_p} v^\dagger (\vec{p}) {\cal P}_\pm^\dagger 
(\vec{p}) \sigma^3 {\cal P}_\pm (\vec{p}) v (\vec{p}) \nonumber \\ 
&=& - \frac{1}{2} \mbox{Tr} \left( {\bf P}_\pm^{(+)} (\vec{p}) 
\sigma^3 \right) \nonumber \\ 
&=& \pm \frac{m_0}{2 E_p} n^3 (\vec{p}) \, . 
\label{tasu2}
\end{eqnarray}
The minus sign on the right-hand side of the first line in Eq. 
(\ref{tasu2}) comes from the fact that the quark obeys the 
Fermi-Dirac statistics. Then, we call the state with ${\cal P}_+ 
(\vec{p}) = 1$ $\left( {\cal P}_+ (\vec{p}) = - 1 \right)$ the 
\lq\lq spin up'' (\lq\lq spin down'') state. 

To see the role of ${\bf P}_s^{(\tau)}$, we first apply 
${\bf P}_s^{(-)} (i \nabla)$ to $q (x)$, 
\begin{eqnarray*}
{\bf P}_s^{(-)} (i \nabla) q (x) & = & \sum_{\bf p} \sum_{s'} 
\sqrt{\frac{m_0}{VE}} \left[ a_{s'} (\vec{p}) {\cal P}_s (\vec{p}) 
\tilde{\Lambda}_- (- \vec{p}) u_{s'} (\vec{p}) e^{- i p \cdot x} 
\right. \nonumber \\ 
&& \left. + b_s^\dagger (\vec{p}) {\cal P}_s (- \vec{p}) 
\tilde{\Lambda}_- (\vec{p}) v_s (\vec{p}) e^{i p \cdot x} \right] 
\, . 
\end{eqnarray*}
Noticing that $\tilde{\Lambda}_- (- \vec{p}) u_{s'} (\vec{p}) = 
u_{s'} (\vec{p})$ and $\tilde{\Lambda}_- (\vec{p}) v_{s'} (\vec{p}) 
= 0$, we have 
\[
{\bf P}_s^{(-)} (i \nabla) q (x) = \sum_{\bf p} 
\sqrt{\frac{m_0}{VE}} a_s (\vec{p}) u_s (\vec{p}) e^{- i p \cdot x} 
\, . 
\]
Thus ${\bf P}_\pm^{(-)} (i \nabla)$ projects out onto the \lq\lq 
spin-up'' (\lq\lq spin-down'') positive energy-state. Similarly, 
using $\tilde{\Lambda}_+ (- \vec{p}) u_{s'} (\vec{p}) = 0$ 
and $\tilde{\Lambda}_+ (\vec{p}) v_{s'} (\vec{p}) = v_{s'} 
(\vec{p})$, we have 
\[
{\bf P}_s^{(+)} (i \nabla) q (x) = \sum_{\bf p} 
\sqrt{\frac{m_0}{VE}} b_s^\dagger (\vec{p}) v_s (\vec{p}) 
e^{i p \cdot x} 
\]
and then ${\bf P}_\pm^{(+)} (i \nabla)$ projects out onto the \lq\lq 
spin-up'' (\lq\lq spin-down'') negative energy-state. 
\setcounter{equation}{1}
\setcounter{section}{1}
\def\theequation{\mbox{\Alph{section}.\arabic{equation}}}
\section{Conservation of $Q_+$ and $Q_-$ up to $O (G_s)$ and $O 
(G_D)$}
In this Appendix, we show that both the \lq\lq magnetic-moment 
(MM)''-polarized  net quark-number charges ($Q_+$ and $Q_-$) and the 
\lq\lq spin''-polarized ones ($Q_+'$ and $Q_-'$) are conserved up to 
and including $O (G_S)$ and $O (G_D)$. 

The Hamiltonian reads 
\begin{eqnarray}
H (x_0) & = & \int d^3 x {\cal H} (x) \equiv H_0 + H_S + H_{PS} + 
H_D \, , \\ 
{\cal H} &=& \bar{q} \left( - i \vec{\gamma} \cdot \nabla + m_0 
\right) q - G_S \left[ \left( \bar{q} q \right)^2 + \left( 
\bar{q} i \gamma_5 \vec{\tau} q \right)^2 \right] \, , \nonumber \\ 
&& - G_D \left[ \left(i \bar{q}^C \epsilon \epsilon^b \gamma_5 q 
\right) \left(i \bar{q} \epsilon \epsilon^b \gamma_5 q^C 
\right) \right] \, . 
\label{Hamiltonian}
\end{eqnarray}
Here $H_0$, $H_S$, $H_{PS}$, and $H_D$ are, in respective order, the 
free-, the scalar quark-antiquark interaction, the pseudoscalar 
quark-antiquark interaction, and the diquark interaction 
Hamiltonians. Since $Q_\pm = Q_\pm^{(-)} + Q_\mp^{(+)}$ and 
$Q_\pm' = Q_\pm^{(-)} + Q_\pm^{(+)}$, it is sufficient to show that 
$\left[ H (x_0), Q_s^{(\tau)} (x_0) \right] = O (G_S^2, G_D^2, G_S 
G_D)$. 

The quark propagator $G$ in the imaginary-time formalism is $G_{11} 
\, \rule[-2mm]{.14mm}{6.0mm} \raisebox{-2.0mm}{\scriptsize{$\; 
|\Delta| = 0$}}$ with $m_0$ for $m$, where $G_{11}$ is as in 
Appendix C: 
\begin{equation}
G = \begin{cases} 
\displaystyle{\sum_{\tau, \, s = \pm} \frac{\gamma_0 {\bf P}_s^{(\tau)} 
(\vec{p})}{p_0 + \tau E_p + \mu_{- \tau s}} \;\;\;\;\;\;\; 
(\mbox{\lq\lq MM''-polarized case}) }\, , \\ 
\displaystyle{\sum_{\tau, \, s = \pm} \frac{\gamma_0 {\bf P}_s^{(\tau)} 
(\vec{p})}{p_0 + \tau E_p + \mu_s} \;\;\;\;\;\;\; 
(\mbox{\lq\lq spin''-polarized case})} \, , 
\end{cases}
\label{denyo}
\end{equation}
where $p_0 = i p_{0 E}$. 
\subsubsection*{Derivation of $\left[ H_0,  \, Q_s^{(\tau)} \right] 
= 0$} 
We write $H_0$ as 
\begin{eqnarray*}
H_0 & = & \int d^{\, 3} x \, \bar{q} \left( - i \vec{\gamma} \cdot 
\nabla + m_0 \right) q \nonumber \\ 
&=& \int d^{\, 3} x \, q^\dagger E \sum_s \left[ {\bf P}_s^{(-)} ( i 
\nabla) - {\bf P}_s^{(+)} ( i \nabla) \right] q \, , 
\end{eqnarray*}
where $E \equiv \sqrt{- \nabla^2 + m^2_0}$. 
Using Eq. (\ref{keypro}), we can easily show that the equal-time 
commutator between $Q_s^{(\tau)}$ and $H_0$ vanishes: 
\[
\left[ Q_s^{(\tau)}, \, H_0 \right] = 0 \, . 
\]
\subsubsection*{Derivation of $\left[ H_{S (PS)}, \, Q_s^{(\tau)} 
\right] = O (G_S^2, G_D^2, G_D G_S)$} 
We start with computing 
\begin{eqnarray} 
\left[ H_{S (PS)} , \, Q_s^{(\tau)} \right] & = & G_S \int d^{\, 3} 
x \left[ \left( \bar{q} \left[ {\bf P}_s^{(\tau)} (i 
\stackrel{\leftarrow}{\nabla}) \Gamma - \Gamma {\bf P}_s^{(\tau)} (i 
\nabla) \right] q \right) \left( \bar{q} \Gamma q \right) \right. 
\nonumber \\ 
&& \left. + \left( \bar{q} \Gamma q \right) \left( \bar{q} \left[ 
{\bf P}_s^{(\tau)} (i \stackrel{\leftarrow}{\nabla}) \Gamma - \Gamma 
{\bf P}_s^{(\tau)} (i \nabla) \right] q \right) \right] \, , 
\label{s1} 
\end{eqnarray} 
where $\Gamma = 1$ for $H_S$ and $\Gamma = i \gamma_5 \vec{\tau}$ 
for $H_{PS}$. 

Let us compute the ensemble average of Eq. (\ref{s1}) up to $O 
(G_S)$ using Eq. (\ref{denyo}): 
\begin{eqnarray} 
\langle \left[ H_{S (PS)} , \, Q_s^{(\tau)} \right] \rangle & = & 
2 G_S V T^2 \sum_{p_0, \, q_0} \int \frac{d^{\, 3} p}{(2 \pi)^3} 
\int \frac{d^{\, 3} q}{(2 \pi)^3} \nonumber \\ 
&& \times \mbox{Tr} \left[ G (\vec{q}) \left\{ \Gamma 
{\bf P}_s^{(\tau)} (- \vec{p}) - {\bf P}_s^{(\tau)} (\vec{q}) \Gamma 
\right\} G (\vec{p}) \Gamma \right] \, , 
\label{s12} 
\end{eqnarray} 
where $V$ is the volume of the system. Appropriate regurarization is 
understood to be introduced. From Eq. (\ref{denyo}), we have 
\begin{equation} 
{\bf P}_s^{(\tau)} (- \vec{p}) G (\vec{p}) = G (\vec{p}) 
{\bf P}_s^{(\tau)} (\vec{p}) \, . 
\label{s13} 
\end{equation} 
Using this relation in Eq. (\ref{s12}), one can readily see that 
the right-hand side vanishes, so that $\left[ H_{S (PS)}, \, 
Q_s^{(\tau)} \right] = O (G_S^2, G_D^2, G_S G_D)$. 
\subsubsection*{Derivation of $\left[ H_D, \, Q_s^{(\tau)} 
\right] = 0 (G_S^2, G_D^2, G_D G_S)$} 
We start with computing 
\begin{eqnarray} 
\left[ H_D , \, Q_s^{(\tau)} \right] & = & 2 G_D \int d^{\, 3} x 
\left[ \left( \bar{q}^C \epsilon \epsilon^b \gamma_5 
{\bf P}_s^{(\tau)} (i \nabla) q \right) \left( \bar{q} \epsilon 
\epsilon^b \gamma_5 q^C \right) \right. \nonumber \\ 
&& \left. - \left( \bar{q}^C \epsilon \epsilon^b \gamma_5 q \right) 
\left( \bar{q} {\bf P}_s^{(\tau)} (i \stackrel{\leftarrow}{\nabla}) 
\epsilon \epsilon^b \gamma_5 q^C \right) \right] \, . 
\label{s14} 
\end{eqnarray} 

Computation of the ensemble average of Eq. (\ref{s14}) up to 
$O (G_D)$ yields 
\begin{eqnarray} 
\langle \left[ H_D , \, Q_s^{(\tau)} \right] \rangle & = & 
16 G_D V T^2 \sum_{p_0, \, q_0} \int \frac{d^{\, 3} p}{(2 \pi)^3} 
\int \frac{d^{\, 3} q}{(2 \pi)^3} \nonumber \\ 
&& \times \mbox{Tr} \left[ G^T (\vec{q}) C \gamma_5 
\left\{ {\bf P}_s^{(\tau)} (- \vec{p}) G (\vec{p}) - G (\vec{p}) 
{\bf P}_s^{(\tau)} (\vec{p}) \right\} \gamma_5 C \right] \, . 
\label{s15} 
\end{eqnarray} 
Using the relation (\ref{s13}) in Eq. (\ref{s15}), one can readily 
see that the right-hand side vanishes, so that $\left[ H_D, \, 
Q_s^{(\tau)} \right] = O (G_S^2, G_D^2, G_S G_D)$. 
\setcounter{equation}{1}
\setcounter{section}{2}
\def\theequation{\mbox{\Alph{section}.\arabic{equation}}}
\section{Propagators} 
In this Appendix, for completeness, we write down the form of 
propagator $G$ with $G^{- 1}$ as in Eq. (\ref{matri}). The 
propagator $G_0$ for the blue quark is obtained from $G$ by taking 
the limit $\Delta \to 0$ and deleting ${\bf 1}_c^\perp$. 
\subsubsection*{Magnetic-moment--polarized quark matter} 
Straightforward manipulation yields, with obvious notation, 
\begin{eqnarray*}
G_{11 (22)} & = & \sum_s \left[ \frac{p_0 - E_p \mp 
\mu_{\pm s}}{(p_0 + E_p \pm \mu_{\mp s}) (p_0 - E_p \mp \mu_{\pm s}) 
- |\Delta|^2} \gamma_0 {\bf P}_s^{(+)} (\vec{p}) \right. \nonumber 
\\ 
&& \left. + \frac{p_0 + E_p \mp \mu_{\mp s}}{(p_0 - E_p \pm 
\mu_{\pm s}) (p_0 + E_p \mp \mu_{\mp s}) - |\Delta|^2} \gamma_0 
{\bf P}_s^{(-)} (\vec{p}) \right] {\bf 1}_f {\bf 1}_c^\perp \, , \\ 
G_{21 (12)} & = & \sum_s \frac{\Delta^\pm}{(p_0 + E_p \pm 
\mu_{\mp s}) (p_0 - E_p \mp \mu_{\pm s}) - |\Delta|^2} 
{\bf P}_s^{(+)} (\vec{p}) \nonumber \\ 
&& + \frac{\Delta^\pm}{(p_0 - E_p \pm \mu_{\pm s}) (p_0 + E_p \mp 
\mu_{\mp s}) - |\Delta|^2} {\bf P}_s^{(-)} (\vec{p}) \nonumber \, , 
\end{eqnarray*}
where $p_0 = i p_{0 E}$. 
\subsubsection*{Spin-polarized quark matter} 
The propagator $G$ is deduced as 
\begin{eqnarray*}
G_{11 (22)} & = & \sum_s \left[ \frac{p_0 - E_p \mp \mu_{- s}}{(p_0 
+ E_p \pm \mu_s) (p_0 - E_p \mp \mu_{- s}) - |\Delta|^2} \gamma_0 
{\bf P}_s^{(+)} (\vec{p}) \right. \nonumber \\ 
&& \left. + \frac{p_0 + E_p \mp \mu_{- s}}{(p_0 - E_p \pm \mu_s) 
(p_0 + E_p \mp \mu_{- s}) - |\Delta|^2} \gamma_0 {\bf P}_s^{(-)} 
(\vec{p}) \right] {\bf 1}_f {\bf 1}_c^\perp \, , \\ 
G_{21 (12)} & = & \sum_s \frac{\Delta^\pm}{(p_0 + E_p \pm \mu_s) 
(p_0 - E_p \mp \mu_{- s}) - |\Delta|^2} {\bf P}_s^{(+)} (\vec{p}) 
\nonumber \\ 
&& + \frac{\Delta^\pm}{(p_0 - E_p \pm \mu_s) (p_0 + E_p \mp 
\mu_{- s}) - |\Delta|^2} {\bf P}_s^{(-)} (\vec{p}) \, . 
\end{eqnarray*}
\setcounter{equation}{1}
\setcounter{section}{3}
\def\theequation{\mbox{\Alph{section}.\arabic{equation}}}
\section{Derivation of Eq. (3.13)} 
In this Appendix, we compute $\mbox{Det} \left( - \beta G^{- 1} 
\right)$ in Eq. (\ref{kihon}) with Eq. (\ref{matri}), 
\[
\mbox{Det} \left( - \beta G^{- 1} \right) = 
\mbox{Det} \left[ - \beta \left( 
\begin{array}{cc} 
\left( G_0^+ \right)^{- 1} {\bf 1}_f {\bf 1}_c^\perp \; & \, 
\Delta^- \\ 
\Delta^+ \; & \, \left( G_0^- \right)^{- 1} {\bf 1}_f 
{\bf 1}_c^\perp 
\end{array} 
\right) \right] \, . 
\]
Using the identities, 
\begin{equation}
\mbox{Det} \left( 
\begin{array}{cc} 
A \; & \, B \\ 
C \; & \, D 
\end{array} 
\right) = (-)^n \mbox{Det} \left( C B - C A C^{- 1} D \right) = 
(-)^n \mbox{Det} \left( B C - B D B^{- 1} A \right) \, , 
\label{kousiki}
\end{equation}
where $A, B, C$, and $D$ are $(n \times n)$ matrices, we obtain 
\begin{eqnarray}
\left[ \mbox{Det} \left( - \beta G^{- 1} \right) \right]^2 & = & 
\mbox{Det} \left\{ \beta^4 \prod_{\tau = \pm} \left[ \Delta^{\tau} 
\Delta^{- \tau} - \Delta^\tau \left( G_0^\tau \right)^{- 1} 
\left( \Delta^\tau \right)^{- 1} \left( G_0^{- \tau} \right)^{- 1} 
\right] \right\} \nonumber \\ 
&=& \mbox{Det} \left\{ \beta^4 \prod_{\tau = \pm} \left[ - 
|\Delta|^2 - \gamma_5 \left( G_0^\tau \right)^{- 1} 
\gamma_5 \left( G_0^{- \tau} \right)^{- 1} \right] {\bf 1}_f 
{\bf 1}_c^\perp \right\} \, , 
\label{c4}
\end{eqnarray}
where use has been made of Eq. (\ref{Delta}). 

Substituting Eq. (\ref{kakuyo}) into Eq. (\ref{c4}), and using Eqs. 
(\ref{D1}), (\ref{D2}), and (\ref{keypro}) we obtain, 
after some algebras, 
\begin{eqnarray*}
\left[ \mbox{Det} \left( - \beta G^{- 1} \right) \right]^2 & = & 
\mbox{Det} \left[ \beta^4 \left\{ F_1 
- 2p_0 \left( p_0^2 - E_p^2 - \mu_+ \mu_- - |\Delta|^2 \right) 
\right. \right. \nonumber \\ 
&& \left. \left. \times \sum_s \left( \mu_s - \mu_{- s} \right) 
\left( {\bf P}_s^{(+)} (- \vec{p}) - {\bf P}_s^{(-)} (- \vec{p}) 
\right) \right\} {\bf 1}_f {\bf 1}_c^\perp \right] \, , 
\end{eqnarray*}
where $p_0 = i p_{0 E}$ (cf. Eq. (\ref{matsu})) and 
\[
F_1 = \left( p_0^2 - E_p^2 - \mu_+ \mu_- - |\Delta|^2 \right)^2 
+ p_0^2 \left( \mu_+ - \mu_- \right)^2 - E_p^2 \left( \mu_+ + \mu_- 
\right)^2 \, . 
\]

Using Eq. (\ref{pro}) with Eqs. (\ref{spinpro}) and 
(\ref{energypro}), and Eq. (\ref{rela}), we obtain, after some 
algebras, 
\[
\left[ \mbox{Det} \left( - \beta G^{- 1} \right) \right]^2 = 
\mbox{Det} \beta^4 \left( 
\begin{array}{cc}
F_{11} \; & F_{12} \\ 
F_{21} \; & F_{22} 
\end{array}
\right) \, . 
\]
Here
\begin{eqnarray*}
F_{11} &=& F_{22} = \left[ F_1 {\bf 1}_2 + \frac{m}{E_p} 
\left( \vec{\sigma} \cdot \vec{n} \right) F_2 \right] 
{\bf 1}_2 {\bf 1}_f {\bf 1}_c^\perp \, , \\ 
F_{12} & = & - F_{21} \equiv - \frac{i}{E_p} \left[ (\vec{p} 
\times \vec{n}) \cdot \vec{\sigma} \, F_2 \right] {\bf 1}_f 
{\bf 1}_c^\perp \, , 
\end{eqnarray*}
where $\vec{n} = \vec{n} (\vec{p})$, ${\bf 1}_2$ is the $(2 \times 
2)$ unit matrix in a spin space, and 
\[
F_2 = 2 p_0 \left( \mu_+ - \mu_- \right) \left( p_0^2 - E_p^2 - 
\mu_+ \mu_- - |\Delta|^2 \right) \, . 
\]

Using Eq. (\ref{kousiki}) again, we obtain, after some algebras, 
\[
\left[ \mbox{Det} \left( - \beta G^{- 1} \right) \right]^2 = 
\mbox{Det} \left\{ \beta^8 \left( F_2^2 - F_1^2 \right) {\bf 1}_2 
{\bf 1}_f {\bf 1}_c^\perp \right\} \, , 
\]
where use has been made of $|\vec{p} \times \vec{n}| = 
\vec{p}^{\, 2} \vec{n}^{\, 2} - (\vec{p} \cdot \vec{n})^2$ and Eq. 
(\ref{rela}). Further manipulation yields (cf. Eq. (\ref{kihon})) 
\[
\ln {\cal Z}_{r, g} = \frac{1}{2} \ln \mbox{Det} \left( - \beta 
G^{- 1} \right) = N_f \sum_n \sum_{\vec{p}} \ln \left( 
\prod_{\rho = \pm} \prod_{\sigma = \pm} \left[ p_0^2 - \left( E_{p, 
\, \rho}^{(\sigma)} \right)^2 \right] \right) 
\]
with 
\begin{equation}
E_{p, \, \rho}^{(\sigma)} = \left[ \left( E_p + \sigma \frac{\mu_+ + 
\mu_-}{2} \right)^2 + |\Delta|^2 \right]^{1 / 2} + \rho \frac{\mu_+ 
- \mu_-}{2} \, . 
\label{finn4}
\end{equation}

Using the relation 
\[
\sum_n \ln \left[ \beta^2 \left( p_0^2 - E^2 \right) \right] = \beta 
\left[ E + 2 T \ln \left( 1 + e^{- \beta E} \right) \right] 
\]
and making the replacement being valid in the large-$V$ limit, 
\[
\sum_n \rightarrow V \int \frac{d^{\, 3} p}{(2 \pi)^3} \, , 
\]
we finally obtain for the contribution from the red- 
and green-quarks to the thermodynamic potential, 
\begin{eqnarray}
\Omega_{r, \, g} & \equiv & - T \frac{\ln {\cal Z}_{r, \, g}}{V} 
\nonumber \\ 
&=& - N_f \int \frac{d^{\, 3} p}{(2 \pi)^3} \sum_{\rho, \, \sigma = 
\pm} \left[ E_{p, \, \rho}^{(\sigma)} + 2 T \ln \left( 1 + 
e^{- \beta E_{p, \, \rho}^{(\sigma)}} \right) \right] \, . 
\label{finn} 
\end{eqnarray}
\end{appendix} 

\end{document}